\documentclass[%
 reprint,
superscriptaddress,
 amsmath,amssymb,
 aps,
]{revtex4-2}

\usepackage{graphicx}
\usepackage{dcolumn}
\usepackage{bm}
\usepackage{hyperref}
\usepackage[mathlines]{lineno}
\usepackage{xcolor}
\usepackage{multirow}
\usepackage{verbatim,upgreek} 

\newcommand{\twC}{$^{\mathrm{12}}$C~}

\newcommand{\siO}{$^{\mathrm{16}}$O~}
\newcommand{\Hy}{$^{\mathrm{1}}$H~}

\newcommand{\Fe}{$^{\mathrm{56}}$Fe~}

\newcommand{\mev}{$~\text{MeV/nucleon}$~}
\newcommand{\detof}{$\Delta \text{E}$-TOF~}

\begin{document}

\preprint{Physical Review C}

\title{Angular differential and elemental fragmentation cross sections of a 400\mev $^{16}\text{O}$ beam on a graphite target with the FOOT experiment}
\author{R. Ridolfi}
\email[Corresponding author: ]{Riccardo.Ridolfi@bo.infn.it}
\affiliation{INFN Section of Bologna, Bologna, Italy}
\affiliation{University of Bologna, Department of Physics and Astronomy, Bologna, Italy}
\author{M. Toppi}
\affiliation{University of Rome La Sapienza, Department of Scienze di Base e Applicate per l'Ingegneria (SBAI), Rome, Italy}
\affiliation{INFN Section of Roma 1, Rome, Italy}
\author{A. Mengarelli}
\affiliation{INFN Section of Bologna, Bologna, Italy}
\author{M. Dondi}
\affiliation{INFN Section of Bologna, Bologna, Italy}
\affiliation{University of Bologna, Department of Physics and Astronomy, Bologna, Italy}
\author{A. Alexandrov}
\affiliation{INFN Section of Napoli, Napoli, Italy}
\author{B. Alpat}
\affiliation{INFN Section of Perugia, Perugia, Italy}
\author{G. Ambrosi}
\affiliation{INFN Section of Perugia, Perugia, Italy}
\author{S. Argir\`o}
\affiliation{University of Torino, Department of Physics, Torino, Italy}
\affiliation{INFN Section of Torino, Torino, Italy}
\author{M. Barbanera}
\affiliation{INFN Section of Perugia, Perugia, Italy}
\author{N. Bartosik}
\affiliation{INFN Section of Torino, Torino, Italy}
\author{G. Battistoni}
\affiliation{INFN Section of Milano, Milano, Italy}
\author{M.G. Bisogni}
\affiliation{University of Pisa, Department of Physics, Pisa, Italy}
\affiliation{INFN Section of Pisa, Pisa, Italy}
\author{V. Boccia}
\affiliation{University of Napoli, Department of Physics "E.~Pancini", Napoli, Italy}
\affiliation{INFN Section of Napoli, Napoli, Italy}
\author{F. Cavanna}
\affiliation{INFN Section of Torino, Torino, Italy}
\author{P. Cerello}
\affiliation{INFN Section of Torino, Torino, Italy}
\author{E. Ciarrocchi}
\affiliation{University of Pisa, Department of Physics, Pisa, Italy}
\affiliation{INFN Section of Pisa, Pisa, Italy}
\author{A. De Gregorio}
\affiliation{University of Rome La Sapienza, Department of Physics, Rome, Italy}
\affiliation{INFN Section of Roma 1, Rome, Italy}
\author{G. De Lellis}
\affiliation{University of Napoli, Department of Physics "E.~Pancini", Napoli, Italy}
\affiliation{INFN Section of Napoli, Napoli, Italy}
\author{A. Di Crescenzo}
\affiliation{University of Napoli, Department of Physics "E.~Pancini", Napoli, Italy}
\affiliation{INFN Section of Napoli, Napoli, Italy}
\author{B. Di Ruzza}
\affiliation{University of Foggia, Foggia, Italy}
\affiliation{INFN Section of Bari, Bari, Italy}
\author{M. Donetti}
\affiliation{CNAO Centro Nazionale di Adroterapia Oncologica, Pavia, Italy}
\affiliation{INFN Section of Torino, Torino, Italy}
\author{Y. Dong}
\affiliation{INFN Section of Milano, Milano, Italy}
\author{M. Durante}
\affiliation{University of Napoli, Department of Physics "E.~Pancini", Napoli, Italy}
\affiliation{Biophysics Department, GSI Helmholtzzentrum f\"ur Schwerionenforschung, Darmstadt, Germany}
\author{R. Faccini}
\affiliation{University of Rome La Sapienza, Department of Physics, Rome, Italy}
\affiliation{INFN Section of Roma 1, Rome, Italy}
\author{V. Ferrero}
\affiliation{INFN Section of Torino, Torino, Italy}
\author{C. Finck}
\affiliation{Universit\'e de Strasbourg, CNRS, IPHC UMR 7871, F-67000 Strasbourg, France}
\author{E. Fiorina}
\affiliation{INFN Section of Torino, Torino, Italy}
\author{M. Francesconi}
\affiliation{INFN Section of Napoli, Napoli, Italy}
\author{M. Franchini}
\affiliation{INFN Section of Bologna, Bologna, Italy}
\affiliation{University of Bologna, Department of Physics and Astronomy, Bologna, Italy}
\author{G. Franciosini}
\affiliation{University of Rome La Sapienza, Department of Scienze di Base e Applicate per l'Ingegneria (SBAI), Rome, Italy}
\affiliation{INFN Section of Roma 1, Rome, Italy}
\author{G. Galati}
\affiliation{University of Bari, Department of Physics, Bari, Italy}
\affiliation{INFN Section of Bari, Bari, Italy}
\author{L. Galli}
\affiliation{INFN Section of Pisa, Pisa, Italy}
\author{M. Ionica}
\affiliation{INFN Section of Perugia, Perugia, Italy}
\author{A. Iuliano}
\affiliation{University of Napoli, Department of Physics "E.~Pancini", Napoli, Italy}
\affiliation{INFN Section of Napoli, Napoli, Italy}
\author{K. Kanxheri}
\affiliation{INFN Section of Perugia, Perugia, Italy}
\affiliation{University of Perugia, Department of Physics and Geology, Perugia, Italy}
\author{A.C. Kraan}
\affiliation{INFN Section of Pisa, Pisa, Italy}
\author{C. La Tessa}
\affiliation{University of Miami, Radiation Oncology, Miami, FL, United States}
\affiliation{Trento Institute for Fundamental Physics and Applications, Istituto Nazionale di Fisica Nucleare (TIFPA-INFN), Trento, Italy}
\author{A. Lauria}
\affiliation{University of Napoli, Department of Physics "E.~Pancini", Napoli, Italy}
\affiliation{INFN Section of Napoli, Napoli, Italy}
\author{E. Lopez Torres}
\affiliation{CEADEN, Centro de Aplicaciones Tecnologicas y Desarrollo Nuclear, Havana, Cuba}
\affiliation{INFN Section of Torino, Torino, Italy}
\author{M. Magi}
\affiliation{University of Rome La Sapienza, Department of Scienze di Base e Applicate per l'Ingegneria (SBAI), Rome, Italy}
\affiliation{INFN Section of Roma 1, Rome, Italy}
\author{A. Manna}
\affiliation{INFN Section of Bologna, Bologna, Italy}
\affiliation{University of Bologna, Department of Physics and Astronomy, Bologna, Italy}
\author{M. Marafini}
\affiliation{Museo Storico della Fisica e Centro Studi e Ricerche Enrico Fermi, Rome, Italy}
\affiliation{INFN Section of Roma 1, Rome, Italy}
\author{M. Massa}
\affiliation{INFN Section of Pisa, Pisa, Italy}
\author{C. Massimi}
\affiliation{INFN Section of Bologna, Bologna, Italy}
\affiliation{University of Bologna, Department of Physics and Astronomy, Bologna, Italy}
\author{I. Mattei}
\affiliation{INFN Section of Milano, Milano, Italy}
\author{A. Mereghetti}
\affiliation{CNAO Centro Nazionale di Adroterapia Oncologica, Pavia, Italy}
\author{T. Minniti}
\affiliation{University of Rome Tor Vergata, Department of Physics, Rome, Italy}
\affiliation{INFN Section of Roma Tor Vergata, Rome, Italy}
\author{A. Moggi}
\affiliation{INFN Section of Pisa, Pisa, Italy}
\author{M.C. Montesi}
\affiliation{INFN Section of Napoli, Napoli, Italy}
\affiliation{University of Napoli, Department of Chemistry, Napoli, Italy}
\author{M.C. Morone}
\affiliation{University of Rome Tor Vergata, Department of Physics, Rome, Italy}
\affiliation{INFN Section of Roma Tor Vergata, Rome, Italy}
\author{M. Morrocchi}
\affiliation{University of Pisa, Department of Physics, Pisa, Italy}
\affiliation{INFN Section of Pisa, Pisa, Italy}
\author{S. Muraro}
\affiliation{INFN Section of Milano, Milano, Italy}
\author{N. Pastrone}
\affiliation{INFN Section of Torino, Torino, Italy}
\author{V. Patera}
\affiliation{University of Rome La Sapienza, Department of Scienze di Base e Applicate per l'Ingegneria (SBAI), Rome, Italy}
\affiliation{INFN Section of Roma 1, Rome, Italy}
\author{F. Peverini}
\affiliation{INFN Section of Perugia, Perugia, Italy}
\affiliation{University of Perugia, Department of Physics and Geology, Perugia, Italy}
\author{F. Pennazio}
\affiliation{INFN Section of Torino, Torino, Italy}
\author{C. Pisanti}
\affiliation{INFN Section of Bologna, Bologna, Italy}
\affiliation{University of Bologna, Department of Physics and Astronomy, Bologna, Italy}
\author{P. Placidi}
\affiliation{INFN Section of Perugia, Perugia, Italy}
\affiliation{University of Perugia, Department of Engineering, Perugia, Italy}
\author{M. Pullia}
\affiliation{CNAO Centro Nazionale di Adroterapia Oncologica, Pavia, Italy}
\author{L. Ramello}
\affiliation{INFN Section of Torino, Torino, Italy}
\affiliation{University of Piemonte Orientale, Department for Sustainable Development and Ecological Transition, Vercelli, Italy}
\author{C. Reidel}
\affiliation{Biophysics Department, GSI Helmholtzzentrum f\"ur Schwerionenforschung, Darmstadt, Germany}
\author{L. Sabatini}
\affiliation{INFN Laboratori Nazionali di Frascati, Frascati, Italy}
\author{L. Salvi}
\affiliation{INFN Section of Perugia, Perugia, Italy}
\affiliation{University of Perugia, Department of Physics and Geology, Perugia, Italy}
\author{C. Sanelli}
\affiliation{INFN Laboratori Nazionali di Frascati, Frascati, Italy}
\author{A. Sarti}
\affiliation{University of Rome La Sapienza, Department of Scienze di Base e Applicate per l'Ingegneria (SBAI), Rome, Italy}
\affiliation{INFN Section of Roma 1, Rome, Italy}
\author{O. Sato}
\affiliation{Nagoya University, Department of Physics, Nagoya, Japan}
\author{S. Savazzi}
\affiliation{CNAO Centro Nazionale di Adroterapia Oncologica, Pavia, Italy}
\author{L. Scavarda}
\affiliation{ALTEC, Aerospace Logistic Technology Engineering Company, Corso Marche 79, 10146 Torino, Italy}
\author{A. Schiavi}
\affiliation{University of Rome La Sapienza, Department of Scienze di Base e Applicate per l'Ingegneria (SBAI), Rome, Italy}
\affiliation{INFN Section of Roma 1, Rome, Italy}
\author{C. Schuy}
\affiliation{Biophysics Department, GSI Helmholtzzentrum f\"ur Schwerionenforschung, Darmstadt, Germany}
\author{E. Scifoni}
\affiliation{Trento Institute for Fundamental Physics and Applications, Istituto Nazionale di Fisica Nucleare (TIFPA-INFN), Trento, Italy}
\author{L. Servoli}
\affiliation{INFN Section of Perugia, Perugia, Italy}
\author{G. Silvestre}
\affiliation{INFN Section of Perugia, Perugia, Italy}
\author{M. Sitta}
\affiliation{INFN Section of Torino, Torino, Italy}
\affiliation{University of Piemonte Orientale, Department of Science and Technological Innovation, Alessandria, Italy}
\author{R. Spighi}
\affiliation{INFN Section of Bologna, Bologna, Italy}
\author{E. Spiriti}
\affiliation{INFN Laboratori Nazionali di Frascati, Frascati, Italy}
\author{L. Testa}
\affiliation{INFN Section of Roma 1, Rome, Italy}
\affiliation{University of Rome La Sapienza, Department of Physics, Rome, Italy}
\author{V. Tioukov}
\affiliation{INFN Section of Napoli, Napoli, Italy}
\author{S. Tomassini}
\affiliation{INFN Laboratori Nazionali di Frascati, Frascati, Italy}
\author{F. Tommasino}
\affiliation{University of Trento, Department of Physics, Trento, Italy}
\affiliation{Trento Institute for Fundamental Physics and Applications, Istituto Nazionale di Fisica Nucleare (TIFPA-INFN), Trento, Italy}
\author{A. Trigilio}
\affiliation{INFN Laboratori Nazionali di Frascati, Frascati, Italy}
\author{G. Traini}
\affiliation{INFN Section of Roma 1, Rome, Italy}
\author{G. Ubaldi}
\affiliation{INFN Section of Bologna, Bologna, Italy}
\affiliation{University of Bologna, Department of Physics and Astronomy, Bologna, Italy}
\author{S. Valentinetti}
\affiliation{INFN Section of Bologna, Bologna, Italy}
\affiliation{University of Bologna, Department of Physics and Astronomy, Bologna, Italy}
\author{A. Valetti}
\affiliation{University of Torino, Department of Physics, Torino, Italy}
\affiliation{INFN Section of Torino, Torino, Italy}
\author{M. Vanstalle}
\affiliation{Universit\'e de Strasbourg, CNRS, IPHC UMR 7871, F-67000 Strasbourg, France}
\author{U. Weber}
\affiliation{Biophysics Department, GSI Helmholtzzentrum f\"ur Schwerionenforschung, Darmstadt, Germany}
\author{R. Zarrella}
\affiliation{INFN Section of Bologna, Bologna, Italy}
\affiliation{University of Bologna, Department of Physics and Astronomy, Bologna, Italy}
\author{A. Zoccoli}
\affiliation{INFN Section of Bologna, Bologna, Italy}
\affiliation{University of Bologna, Department of Physics and Astronomy, Bologna, Italy}
\author{M. Villa}
\affiliation{INFN Section of Bologna, Bologna, Italy}
\affiliation{University of Bologna, Department of Physics and Astronomy, Bologna, Italy}
\collaboration{The FOOT Collaboration}
\begin{abstract}
This paper presents the measurements of the angular differential cross sections for the forward production of He, Li, Be, B, C and N nuclei in the fragmentation process of a 400\mev \siO beam interacting with a graphite target. Due to the limited data available in this energy regime, these measurements of nuclear fragmentation cross sections are relevant to improve nuclear interaction models for Particle Therapy and space radioprotection applications. The data analyzed in this paper were collected during a measurement campaign carried out at the GSI Helmholtz Center for Heavy Ion Research facility in Darmstadt (Germany) by the FOOT collaboration. The results are compared with similar results found in the literature and with a previous FOOT measurement of the same process, using the same setup, from a previous pilot run performed at GSI. The pilot run data, however, had limited statistics and only allowed for the measurement of elemental fragmentation cross sections integrated in the setup acceptance.
This data set, with statistics more than 100 times larger compared to the data collected in the previous run, enabled the measurement of angular differential cross sections, fully exploiting the granularity of the FOOT \detof system.
Furthermore, a better comprehension of the FOOT apparatus allowed to improve the analysis techniques, leading to a reduction in the final systematic uncertainties. 
The cross section results have been compared with some of the prominent Monte Carlo models of FLUKA and Geant4 dedicated to the energy range of interest for light ion fragmentation physics.
\end{abstract}

\maketitle

\section{Introduction}
Nuclear fragmentation cross section measurements are crucial for advancing in several
fields, including Particle Therapy (PT), 
Radioprotection in Space (RPS) and nuclear structure studies~\cite{Durante2016,  Durante2011}. For instance, in PT understanding the fragmentation of ion beams as they interact with human tissue can potentially improve cancer treatments, as it helps in accurately predicting the dose distribution and minimizing damage to surrounding healthy tissues~\cite{Kramer2010,Thwaites2013}. Moreover, an accurate description of fragmentation phenomena can also shed light on the biological effectiveness in proton therapy~\cite{Tommasino2015}.

Similarly, in RPS these measurements would be of great importance for 
assessing the risks posed by cosmic radiation to astronauts, as they help in developing 
effective shielding strategies~\cite{Wilson1997}. Indeed, space radiation constitutes one of the major risks for space exploration beyond Low Earth Orbit (LEO) which is
among future plans of several national space agencies and private companies~\cite{Heinbockel2011}. Radiation hazards could be so important to prevent deep space missions due to huge costs and unacceptable risks for the astronauts given the lack of effective countermeasures so far~\cite{Durante2014}.

These fields share a common ground both for ions involved (ranging from \Hy to \Fe with a focus on ions with nuclear charge number $Z \leq 8$) and kinetic energies $100-1000$\mev (depending on the ion).
However, the phenomena at play are known with poor accuracy due to the lack or the poor precision of the relevant fragmentation cross section measurements~\cite{Norbury2012,Luoni2021,Norbury2020}.
This translates in a poor precision in the computation of the biological dose due to the fragments with respect to the one required for both PT and RPS applications~\cite{Sommerer2006,Sato2009}.
The measurement of the missing fragmentation cross sections would allow to benchmark and update the existing nuclear models implemented in Monte Carlo (MC) and deterministic codes, used for the dose computation in both the PT and RPS applications~\cite{Bohlen2010,Dodouet2014}.
Double differential cross sections in the fragment production angle and kinetic energy would be of great value to this purpose. Despite recent progress, only a small number of beam-target-energy combinations have been explored~\cite{Dodouet2013,Dodouet201406,Pleskac2012,Toppi2016,Zeitlin2011,Webber90,Mattei2020}.

The FOOT experiment~\cite{frontiersFOOT} has been designed to address this data deficit by measuring fragmentation cross sections
in the nuclear interactions between ion beams (such as protons, Helium, 
Carbon, and Oxygen)
and targets of interest for PT, like H, C and O, which are the most abundant elements in tissues. It also includes targets of interest for shielding in RPS like hydrogen-enriched targets~\cite{frontiersFOOT}.
FOOT is a fixed target experiment whose setup includes two complementary configurations: an electronic setup
with a magnetic spectrometer and charge/mass identification capabilities, for measuring forward emitted fragments with Z~$\ge$~2, and an emulsion spectrometer for higher angular acceptance measurements of fragments with Z~$\le$~3. Details of the FOOT experiment design and some preliminary results can be found in~\cite{frontiersFOOT,FOOTGSI2019,Galati2024}.
This study analyzes data acquired at the GSI Helmholtz Center for Heavy Ion Research facility in Darmstadt in 2021 with the electronic setup.
At that time only a part of the final FOOT detector, as described in detail in~\cite{frontiersFOOT}, was operational. 
The setup, consisting of a detector for the beam monitoring and a system for the Time-of-Flight and the energy loss measurements, was used to identify the charge Z of the fragments and to measure their emission angle allowing to perform elemental cross section measurements.
The same setup was used in a previous data acquisition campaign in 2019 in GSI, described in detail in~\cite{FOOTGSI2019}.

This paper presents the measurement of the angular differential cross sections for
the forward production of $2\leq Z \leq 7$ nuclei
in the fragmentation process of a 400\mev \siO beam interacting with a graphite
target.
In terms of statistics and results, this work extends and supersedes the 2019 study~\cite{FOOTGSI2019}, where a limited statistics allowed to measure only elemental fragmentation cross section integrated in the full geometrical acceptance of the setup. 
The present work provided more than a factor of 100 improvement in statistics compared to the previous campaign which allowed measurement of the angular differential cross sections for charged fragments with FOOT. In Sec.~\ref{sec:materials} the FOOT setup used in this analysis, its performance and the data collected are described.
In Sec.~\ref{sec:analysis} the analysis strategy to measure cross sections is discussed.
Systematic uncertainties from detector effects and the analysis method are also assessed.
In Sec.~\ref{sec:results} the results of the paper are discussed taking into account previous studies and earlier works of the FOOT experiment~\cite{FOOTGSI2019} with a focus on the improvements enabled by higher statistics and enhanced analysis techniques while in Sec.~\ref{sec:MCcomparison} a comparison between experimental results and predictions of Monte Carlo models in FLUKA and Geant4 is presented.

\section{Material and methods}
\label{sec:materials}
\subsection{Experimental setup}
\label{sec:experimentalsetup}
\begin{figure}[h]
\centering
\includegraphics[width=0.47\textwidth]{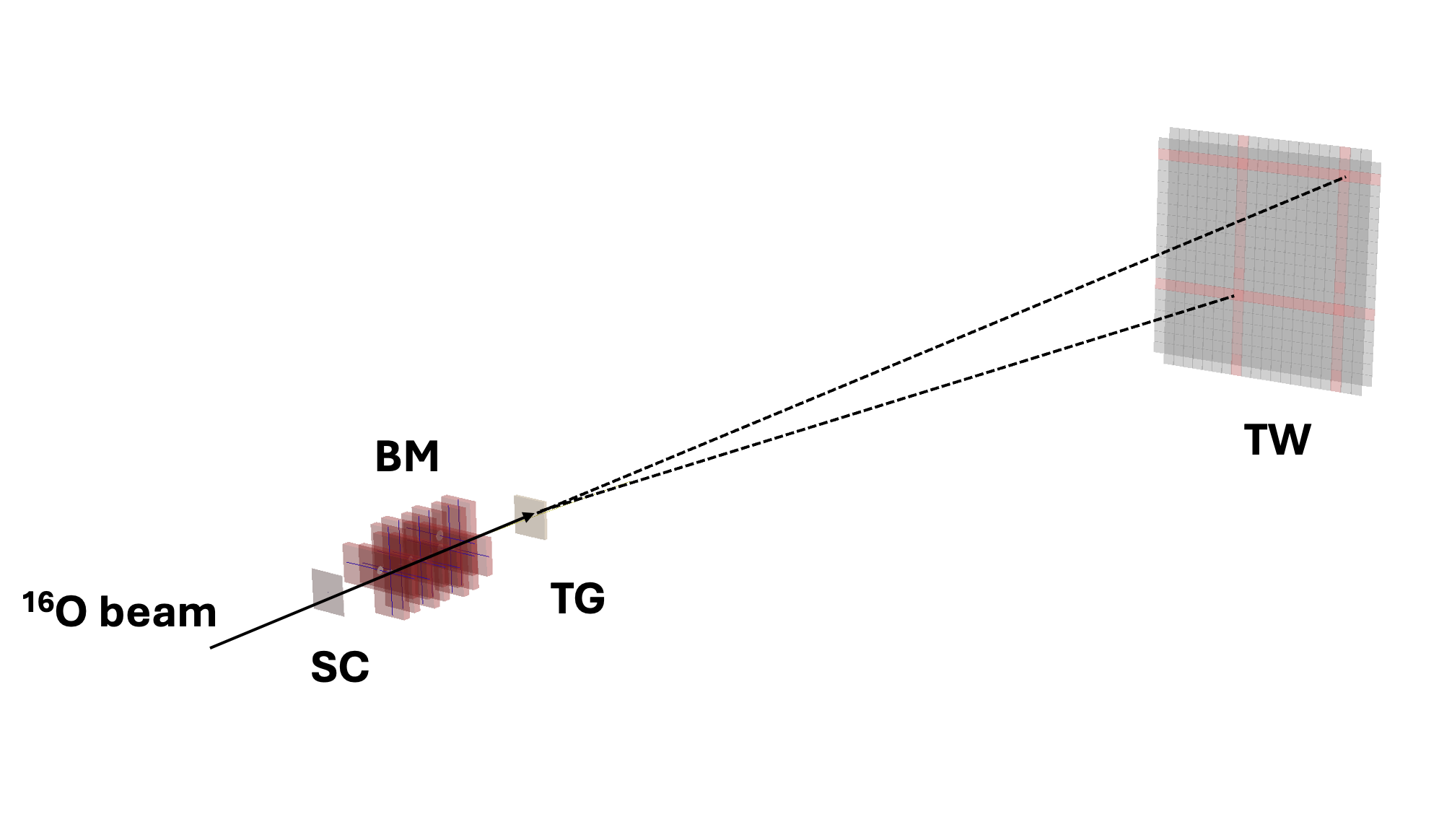}
\caption{Schematic view of the GSI experimental setup. 
The $^{16}$O beam passes through the Start Counter and the Beam Monitor, before impinging on the 5 mm thick graphite target. The produced fragments emitted with a polar angle $\leq 5.7 ^{\circ}$ can be identified by the TOF Wall detector, about 193 cm downstream of the target~\cite{FOOTGSI2019}.}
\label{fig:gsi}
\end{figure}

The experimental setup used in the data campaign analyzed in this paper is identical to the one used in the GSI 2019 data taking campaign and described in detail in~\cite{FOOTGSI2019} (see Figure~\ref{fig:gsi}). The fragmentation due to the interaction of a \siO beam of 400\mev with a 5~mm graphite target (TG) is studied.
Two detectors, upstream the graphite target, the Start Counter (SC) and the Beam Monitor (BM) measure the incoming Oxygen ions. Their design minimizes the pre-target overall material budget in order to keep the fragmentation interactions inside themselves at the percentage level with respect to the one in the target.
Downstream of the target, the only detector is the TOF Wall (TW)
able to provide identification of the charge Z of the fragments~\cite{FOOTGSI2019}.
The FOOT detector was installed in the Cave-A (HTA) of the GSI facility. The FWHM beam size was (0.5,0.4)~cm in $(X,Y)$ and the beam intensity was kept in the interval (0.5-1.0)~kHz in order to keep the pile-up from Oxygen ions to a sustainable rate for the FOOT detectors.

The Start Counter~\cite{Traini2020} is a 250~$\upmu$m thick plastic scintillator (EJ-228), placed at the very beginning of the setup, which measures the number of the incoming \siO ions and provides the start of the time of flight (TOF) measurement with a resolution of $\sim$~70~ps~\cite{Kraan2021}. 
The Beam Monitor is a drift chamber consisting of twelve wire layers, with X-Y views. With a tracking efficiency higher than $90\%$ and a lower limit on the spatial resolution of $60\,\upmu\text{m}$~\cite{Dong2021}, it provides the measurement of the direction and the interacting point of the beam ions on the target. The distances between SC, BM and TG have been minimized as much as possible. This choice minimizes the effects of the multiple scattering and maximizes the position resolution on the BM track projection on the TG.

The TOF Wall detector is composed of two layers of $20$ plastic scintillator bars (EJ-200) each, arranged orthogonally to provide X-Y views. This setup forms a $40\times 40\,\text{cm}^2$ active area detector providing the measurements of the energy deposited $\Delta$E, with a resolution of $\sigma(\Delta E)/\Delta E\sim 5$~\%, the TOF stop, with a resolution of $\sim$~20~ps for \siO ions~\cite{Kraan2021}, and the hit position with the granularity provided by the bar crossing dimension of 2$\times$2 cm$^2$. 
To ensure that most fragments (except protons and neutrons) are fully measured, the TW is positioned approximately 193~cm downstream of the target. MC simulations and prior measurements~\cite{frontiersFOOT,Dodouet2014} showed that this arrangement maximizes the detector's coverage for He fragments emitted within a maximum angle of 10$^\circ$ and consequently also for heavier ions.
The position of the TW was chosen to minimize multiple hits within the fixed granularity. According to a MC simulations the chosen granularity keeps the pile-up of multiple fragments in the same bars cross below $1\%$~\cite{FOOTGSI2019,frontiersFOOT}.
The TW detector was not optimized for the detection of neutrons and protons and these products were not analyzed.
The TW dimension and distances from the target set the geometrical acceptance of the setup. Taking into account also a 1~cm shift between the center of the upstream region (SC, BM and TG) with respect to the one of the TW, the polar angle acceptance for this analysis is $\theta \le 5.7^\circ$.

\subsubsection*{Charge Z identification}
\label{sec:ZID}
\begin{figure}
\includegraphics[width=0.47\textwidth]{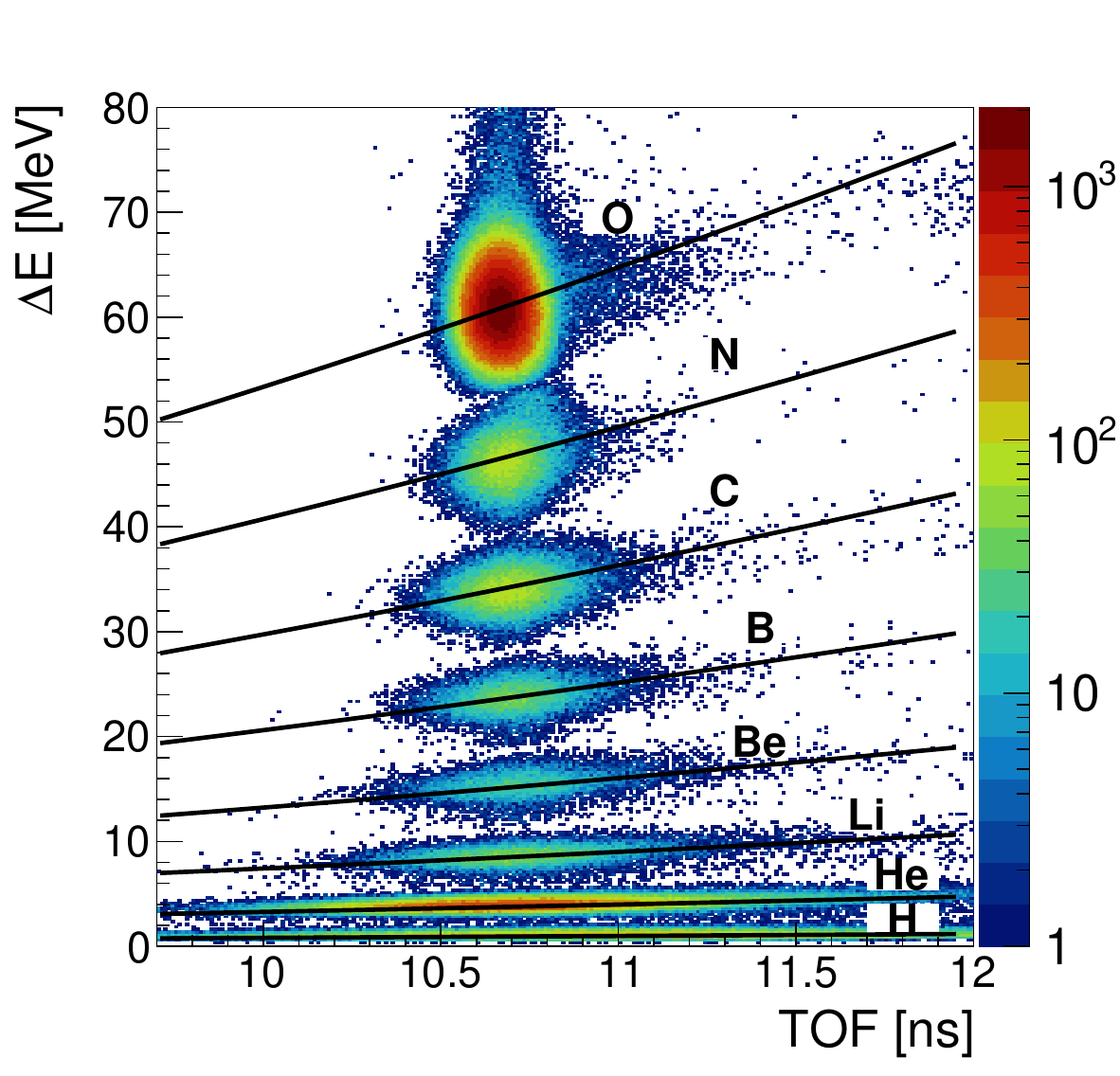}
\caption{$\Delta$E vs TOF distribution for the data collected at GSI in 2021, analyzed in this work. The Bethe-Bloch curves for Z=1 (Hydrogen, lower curve) up to Z=8 (Oxygen, higher curve) are superimposed.
The separation between the individual fragment charges (Z) is clearly visible.}
\label{fig:ZID}
\end{figure}

The simultaneous measurement of the $\Delta$E in each TW bar and the TOF between that bar and the SC allowed the identification of the nuclear charge (Z) of that ion.
As detailed in \cite{FOOTGSI2019}, from a parametrization with a Bethe-Bloch curve of the $\Delta$E as a function of TOF, for each TW layer, the charge Z of each fragment was extracted. 
A fragment crossing the TW  hits a pair of X and Y bars, each with an assigned Z coordinate, thus identifying the point of passage of the fragment. The two bars are associated to the same fragment and clusterized in a TW point, with a position resolution provided by the bar crossing of 2$\times$2~cm$^2$. In order to have a clean Z identification, only pairs of X-Y bars sharing the same reconstructed Z are selected and clusterized in a TW point~\cite{FOOTGSI2019}.
The fragment hit position along the bar, extracted with the time difference measured at both the edges of a single bar, is used to improve the clusterization of the X-Y bars forming a TW point~\cite{FOOTGSI2019}. This is fundamental whenever more than one fragment hits the TW in the same event.
The $\Delta$E and TOF assigned to each TW point is the average between the ones of the two bars forming it. In order to show the separation of the different fragments charge Z in Fig.~\ref{fig:ZID} the Bethe-Bloch curves used for the charge Z identification (ZID) of the fragment are superimposed to the distribution of $\Delta$E vs the TOF for each reconstructed TW point.
        
\subsection{Data Sample and Trigger Strategy}
The data analyzed in this paper were collected during the FOOT data taking campaign at GSI in 2021, with the setup reported in Sec.~\ref{sec:experimentalsetup}. Two data samples were collected: one with a \siO beam of 400\mev impinging on a $5\,\text{mm}$ carbon target and another acquired without target.
The second sample contains only fragmentation coming from the interaction of the beam with the air and the FOOT setup and it is used to evaluate the background in the cross section analysis (see Sec.~\ref{sec:analysis}).

The data sample with the carbon target was collected using two different trigger strategies: the Minimum Bias (MB) trigger was issued whenever a primary ion of \siO goes through the SC, while the fragmentation trigger (FRAG) rejects most of the primaries events reaching the TW,
and is issued in presence of TW hits due to fragments of interest for cross section measurement. Details of the FOOT trigger strategy are discussed in~\cite{Galli2023}. About 1.2$\times 10^6$ events were collected with the MB trigger implemented requiring in the SC a majority of 5 of its readout channels over the total of 8~\cite{Traini2020}. About 2$\times 10^6$ events were acquired with the FRAG trigger, able to reject most of the \siO reaching the TW, requiring a coincidence of a MB trigger with a veto from events with an energy release in the TW central bars compatible with a primary ion of the beam.
As detailed in~\cite{Galli2023}, the veto is implemented setting amplitude thresholds on the TW central bars, tuned during MB runs. The thresholds were chosen as a trade off between the rejection of non-fragmentation events (Oxygen) while keeping most of Nitrogen and lower Z fragments which can be affected by a too tight choice of the thresholds.
The FRAG trigger was designed to mostly reject unreacted \siO ions but some fragments might have been rejected.
Thanks to the loose thresholds used during the acquisition of events with the FRAG trigger, this effect was estimated to be negligible. 
Counting the number of times the FRAG trigger conditions are valid during MB runs the ratio between the number of events labeled as FRAG events and their total number provides the FRAG \emph{trigger acceptance factor} $\varepsilon_{\text{trig}}^{\text{FRAG}}$ which resulted to be $15.99\pm0.03\%$. 

All the FRAG events were acquired along with a fraction of MB events: this means that the data sample contains both FRAG events and MB events (1 out of 10 in this case).
About 5.7$\times 10^4$ events were acquired with the MB trigger for the sample without target for background studies.

\subsection{Monte Carlo simulation}
\label{sec:mc}
A detailed Monte Carlo (MC) simulation of the setup described in Sec.~\ref{sec:experimentalsetup} was carried out using the FLUKA MC code \cite{Battistoni2016}. The MC simulation was tailored to reflect the detector geometry, the passive materials and the alignment adjustments, including shifts and rotations of the setup~\cite{Dong2024}. Both the setups with and without the target were simulated. The \siO beam position and its transverse size were extracted from data and implemented in the MC simulation.
Simulation outputs were processed by the FOOT reconstruction software performing the reconstruction as in data. The reconstruction of the quantities of interest, tracks for BM and points for TW and their performances were already studied and optimized in MC in other works~\cite{Dong2021,FOOTGSI2019}.
Detector responses, including spatial resolution in BM and energy loss and TOF resolutions in SC and TW, were extracted from data and dedicated calibration runs, as described in~\cite{Dong2021,FOOTGSI2019,Morrocchi2019,Traini2020}.  Subsequently this information has been used in MC applying a Gaussian smearing of the quantities computed by FLUKA.
The tuning of MC on data is fundamental to make the MC reliable for what concerns the fragment Z identification and the angle of the reconstructed fragments~\cite{FOOTGSI2019}, which are the two quantities needed to extract the fragments raw yields for the angular differential cross sections, as discussed in the next section~\ref{sec:strategy}.

\section{Data Analysis}
\label{sec:analysis}
\subsection{Analysis Strategy}
\label{sec:strategy}
The goal of this analysis is to evaluate the angular differential cross sections for the fragmentation process of a \siO beam of 400\mev interacting with a graphite target
for the forward production ($\theta\leq 5.7 ^\circ$) of He, Li, Be, B, C and N nuclei.
The elemental cross section for each fragment charge $Z$ is computed as follow:
\begin{equation}\label{eq:XSangle}
\dfrac{\text{d}\sigma}{\text{d}\Omega}(Z,\theta)=\dfrac{Y(Z,\theta)}{N_{\text{prim}} \cdot N_{\text{TG}}\cdot\varepsilon(Z,\theta)\cdot\Delta\Omega}
\end{equation}
where $Y(Z, \theta)$ is the selected number of fragments of a given
charge Z measured by TW at a given angle $\theta$, $N_\text{prim}$ is the selected number of primaries impinging
on the target, $\varepsilon (Z, \theta)$ is the
efficiency for a given charge in a given angle computed using the MC simulation (see Sec.~\ref{sec:mc_corr} and~\ref{sec:eff}), $\Delta \Omega$ is the solid angle bin width and $N_{\text{TG}}$ is the number of interaction centers in the target per unit surface which can be written as:
\begin{equation}
N_{\text{TG}} = \dfrac{\rho \cdot d\cdot N_{\text{A}}}{A}
\end{equation}
where $\rho = 1.83\,\text{g/cm}^3$ is the graphite target density, $d = 0.5\,\text{cm}$
is the target thickness, $N_{\text{A}}$ is the Avogadro number and $A=12.01\,\text{g/mol}$ 
is the graphite mass number.
Details on how the factors entering in Eq.\ref{eq:XSangle} are computed or measured are given below.

In order to measure the fragmentation yields $Y(Z, \theta)$ produced in the target, the target-out fragmentation, mostly produced by the interaction of the \siO beam with the air and the FOOT setup, has to be taken into account.
This contribution has to be considered as background with respect to the in-target fragmentation and need to be detected, identified and removed.
The reconstruction workflow is the same for the samples with and without target: firstly, the angle of the primary particle and its impact point on the target are calculated by the BM. Then, the angle between the impact point on target and the reconstructed point on the TW is calculated, and the charge Z of the fragment is identified by the TW.
Since the TW cannot distinguish whether fragments come from the target or not,
all the valid points are taken into account and their emission angle is calculated as pointed out
above even if fragments are not generated in the target. The background which arises from these wrong reconstructions is
removed using the sample without target.
In conclusion the quantity $Y(Z,\theta)/N_{\text{prim}}$ of Eq.~\ref{eq:XSangle} can be rewritten as 
\begin{equation} \label{eq:YieldSub}
\frac{Y(Z,\theta)}{N_{\text{prim}}} = \frac{Y_{\text{TG}}(Z,\theta)}{N_{\text{prim,TG}}} - \frac{Y_{\text{noTG}}(Z,\theta)}{N_{\text{prim,noTG}}}
\end{equation}
where the subscript TG (noTG) refers to the runs with (without) the target.
In this way, fragment yields are correctly subtracted taking into account the different number of primaries in the two data samples.

The number of primaries N$_\text{prim}$, in MB runs, corresponds to the number of MB trigger
$N_{\text{trig}}^{\text{MB}}$ surviving the selection cuts for pile-up removal on the SC and on the BM. Multiple primary particles in the same event (i.e. pile-up events) are removed looking both at the SC raw signal and at the number of BM reconstructed tracks.
The cleaning of the pile up events in the SC and the request of events with only one BM track reduces the number of MB trigger to be considered for the analysis of 10-12\% depending on the instant beam rate (which can rise up to 50~kHz).
To compute the number of primaries N$_\text{prim}$, in FRAG runs, the number of FRAG trigger N$_\text{trig}^\text{FRAG}$, surviving the same pile-up selections in SC and BM, has to be further divided by the trigger acceptance factor, in order to consider the effect of the trigger selection:
\begin{equation} \label{eq:FragTrig}
N_{\text{prim}}= N_{\text{trig}}^{\text{FRAG}}/\varepsilon_{\text{trig}}^{\text{FRAG}}
\end{equation}

Pile-up selections in BM and SC are performed before the effective beam arrival on the target, thus no significant biases are introduced in the analysis. After the selection cuts $\approx 1.7\times10^6$ MB events, considering both MB runs and MB events in FRAG runs, and $\approx 6.8\times10^5$ FRAG events (corresponding to more than four million primary particles) were used for the analysis. For the background only $\approx 5\times10^4$ were selected due to the limited amount of statistics without 
target. Thus, the error on cross section measurements, especially on angular ones, was mainly driven by the uncertainty in the number of fragments measured in the sample without target.
The background subtraction at yields level, in Eq.~\ref{eq:YieldSub}, is performed after the calculation of 
$Y_{\text{TG}}/N_{\text{prim,TG}}$
as weighted average of MB and FRAG data samples.

The minimum angular bin width used in the cross section (Eq.\ref{eq:XSangle}) was set taking into account the TW granularity ($\approx 0.6^{\circ}$ at the target position) while the choice of the number of bins was mainly driven by the available statistics in background data, which dominate the statistical error.
The fragmentation cross sections are evaluated in the geometrical acceptance of the apparatus of $\theta\le 5.7^\circ$.
Integrating the Eq.~\ref{eq:XSangle} over the solid angle covered, the elemental cross section for each charge $Z$ can be extracted as follows:
\begin{multline}
\sigma (Z) = \int _{\Omega}\dfrac{\partial \sigma}{\partial \Omega} \text{d}\Omega = \dfrac{Y(Z)}{N_{\text{prim}} \cdot N_{\text{TG}} \cdot \varepsilon (Z)}
\label{eq:XStotal}    
\end{multline}
In the present work both the integral and the angular differential cross sections have been measured and are discussed in the results (see Sec.~\ref{sec:results}).

\subsection{MC corrections to the cross sections}
\label{sec:mc_corr}
The MC simulation (Sec.~\ref{sec:mc}) has been used to compute the efficiencies and the corrections needed in the measurement of the fragmentation cross sections (Eq.~\ref{eq:XSangle} and~\ref{eq:XStotal}) and to verify the analysis strategy discussed in the previous Sec.~\ref{sec:strategy}.
A purity correction, discussed in Sec.~\ref{sec:purity}, is introduced to correct the fragmentation yield $Y(Z,\theta)$ to take into account the mis-identification of the Z charge due to the ZID algorithm of TW detector.
The efficiency, detailed in Sec.~\ref{sec:eff}, is introduced to correct for the missing fragments, produced in the fragmentation process, but lost in reconstruction.
The analysis strategy for the cross section evaluation, shown in the previous Sec.~\ref{sec:strategy}, was validated against the MC simulation: the raw yields reconstructed by the TW have been measured from the subtraction between samples with and without target and corrected by purity and efficiency MC corrections. The obtained MC fragmentation cross sections, differential in angle and for each fragment atomic charge Z, were compared to the true MC production cross sections. The true MC production cross sections were obtained by selecting fragment tracks with a given Z$_\text{true}$ and emission angle $\theta_\text{true}$, produced in the target within the TW detector acceptance ($\theta_\text{true} < \text{5.7}^\circ$). A residual discrepancy between reconstructed and true MC cross sections has been observed and attributed to the migration between angular bins caused by the TW limited granularity.
This discrepancy was corrected with an angular unfolding procedure~\cite{Schmitt17} in the analysis (see Sec.~\ref{sec:unfolding}), which allowed to dramatically improve the agreement between reconstructed and true MC cross section, expecially in the case of Carbon (Z=6) and Nitrogen (Z=7) fragments for which a narrow angular distribution is expected~\cite{frontiersFOOT}.
The validation of the analysis strategy in the MC, including background statistical subtraction between samples with and without target, efficiency and purity corrections to the fragmentation yields and finally angular unfolding procedure to cure the bin migration, confirmed the possibility to apply the same strategy in data. In particular after the background subtraction, the purity correction is applied to remedy for the Z charge mis-identification, then the unfolding procedure is used to solve for the angular bin migration and finally the efficiency correction is applied to account for the not reconstructed fragments. The cross section results are shown in Sec.~\ref{sec:results}.

\subsubsection{Purity}
\label{sec:purity}
In this analysis, the TW is the only detector able to identify the charge of the fragments using the energy release $\Delta$E in a TW bar and the TOF between SC and TW. The purity is a quantity which is related to the ZID algorithm performances of the TW and depends on the energy loss and TOF resolutions, which in MC are tuned from data, as explained in Sec.~\ref{sec:mc}~\cite{FOOTGSI2019}. To account for the charge Z mis-identification it was needed to introduce a purity correction, which is calculated in MC for each charge and angle, and can be written as:
\begin{equation}
    P(Z,\theta) =\dfrac{N(Z_\text{reco}=Z_\text{true},\theta_\text{reco})}{N(Z_\text{reco},\theta_\text{reco})}
    \label{eq:purity}
\end{equation}
where $N(Z_\text{reco},\theta_\text{reco})$ is the number of fragments reconstructed by the TW
with charge $Z_\text{reco}$ while $N(Z_\text{reco}=Z_\text{true},\theta_\text{reco})$ is the number of fragments reconstructed by the TW
with charge $Z_\text{reco}$ equal to the true charge $Z_\text{true}$ from MC.
The purity correction is applied as a multiplying factor to the fragmentation yield: 
\begin{equation} \label{eq:purity2}
Y(Z,\theta) = Y_{\text{raw}}(Z,\theta)\cdot P(Z,\theta)
\end{equation}
where $Y_{\text{raw}}(Z,\theta)$ is the yield subtracted of Eq.~\ref{eq:YieldSub}, before the purity correction application. The obtained $Y(Z,\theta)$ is the one entering the Eq.~\ref{eq:XSangle}.
The purity correction is found to be more than 90\% for all the fragment charge Z, with the exception of the case of the Li, for which the purity goes down to the 70\%. This drop is due to the contamination of events with a pair of He fragments entering in a single TW bar cross which release an energy $\Delta$E comparable to the one of a Li, causing a mis-identification of the two He ions in a Li. Two He fragments emitted in a narrow angle is a well known process, in fragmentation physics at these energies, see for example~\cite{Zeitlin2011}, and for the FOOT setup of this analysis it is an unavoidable contamination for Li fragments.

\subsubsection{Unfolding procedure}
\label{sec:unfolding}
As already pointed out, the detector effects on angle measurement have to be taken into account when
dealing with angular differential cross sections. In particular, Multiple Coulomb Scattering and TW granularity
play a major role.
To evaluate these effects, the matrix $\theta_{\text{true}}$~vs~$\theta_{\text{reco}}$ is built (in the following the \emph{response matrix}), where $\theta_{\text{reco}}$ is the measured scattering angle with respect to the original Oxygen beam direction, while $\theta_{\text{true}}$ is the production angle
of fragments born in the target (the scattering inside the target resulted to be negligible).

The unfolding procedure~\cite{Schmitt17} is based on a Bayesian iterative algorithm \cite{dagostini2010improved} \cite{DAGOSTINI1995487} as implemented in RooUnfold \cite{adye2011unfolding}. 
In the unfolding of binned data,
the effects of the acceptance and resolution are expressed in terms of a two-dimensional
response matrix, $C_{ij}$, where each element corresponds to the probability of an event in the $i$-th generator-level bin being reconstructed in the $j$-th measurement bin. The unfolding algorithm combines the 
measured spectrum with the response matrix to form a likelihood, takes as input a prior for the 
specific kinematic variable and iterates using the posterior distribution as prior for the next 
iteration. The MC distribution is used as the initial prior and three iterations are 
performed. The number of iterations is optimized to balance the unfolding stability with respect to 
the previous
iteration and the growth of the statistical uncertainty. The final choice of three iterations is driven by the minimization of the average correlation factor~\cite{Schmitt17}.

\begin{figure}
\includegraphics[width=0.47\textwidth]{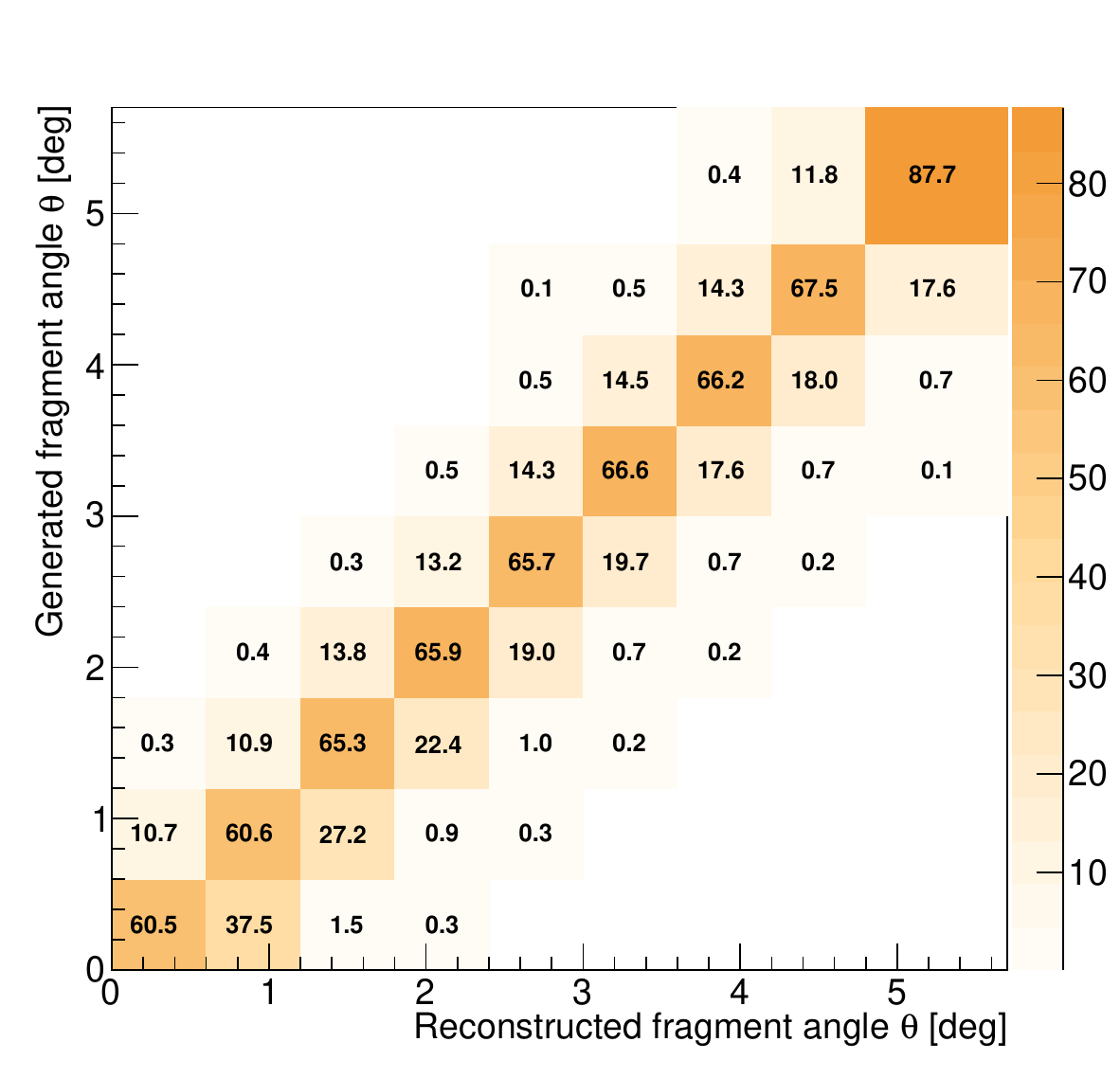}
\caption{Normalized response matrix of the He fragments. Only bins where the migration is greater than $0.1$\% are shown.}
\label{fig:Response}
\end{figure}

The response matrix for the angular distribution of the He fragments is taken as an example of the impact of the unfolding correction and shown in Fig.~\ref{fig:Response}.
In order to ensure that the MC sample used for the unfolding training is not introducing a bias on 
the data measurement, a study is performed by changing the pseudo-data distribution. The MC angular spectrum is re-weighted using continuous functions to alter the final shape with the same binning. The study confirms that the altered shapes unfolded based on the nominal corrections are preserved within statistical uncertainties.     

\subsubsection{Efficiency}
\label{sec:eff}
\begin{figure}
    \includegraphics[width=0.47\textwidth]{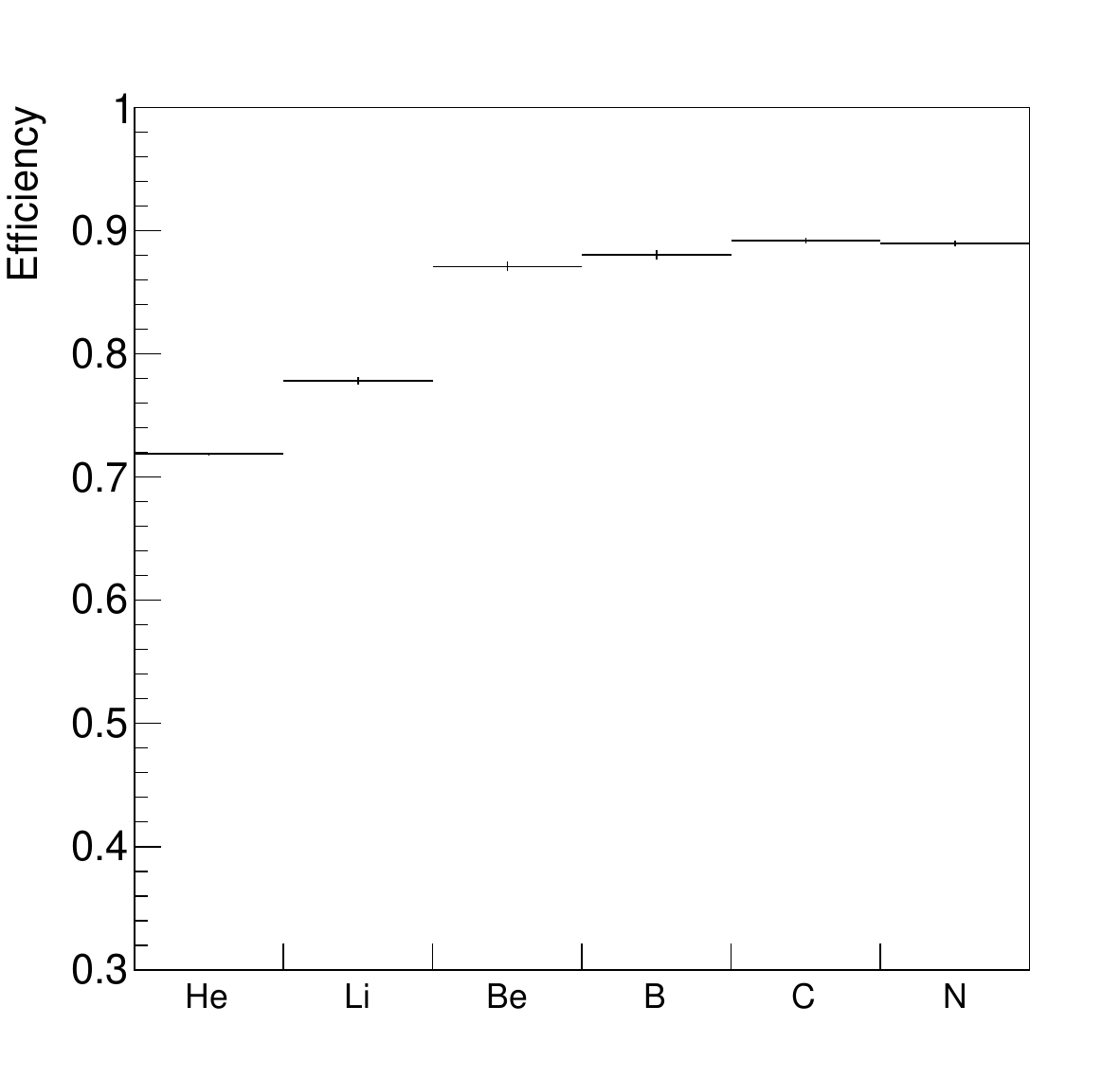}
    \caption{Efficiencies for each reconstructed fragment of charge Z emitted in the TW acceptance of $\text{5.7}^\circ$.}
    \label{fig:eff}
\end{figure}

The efficiency for each Z and for each angle $\theta$ is computed in order to account for the fragments not reconstructed by the TW detector.
Also fragments produced in the TG, within the TW detector acceptance ($\theta < \text{5.7}^\circ$), but not reaching the TW due to multiple coulomb scattering or fragmentation in air, are contributing to the efficiency computation.
In particular, while this last contribution is almost negligible, the impact of the TW reconstruction which discards association of X-Y bars with different Z charge reconstructed, in order to maximize the purity of the charge Z of the TW points, has an important impact~\cite{FOOTGSI2019}.
The efficiency is defined for each fragment of charge Z$_\text{true}$ and for each angle $\theta_\text{true}$ as follows:
\begin{equation}
\varepsilon(Z,\theta) = \dfrac{N_\text{TW}(Z,\theta)}{N_\text{prod}(Z,\theta)}
\end{equation}
where $N_\text{TW}(Z,\theta)$ are the fragments with charge $Z_\text{true}$ and emission angle $\theta_\text{true}$ reconstructed by the TW and $N_\text{prod}(Z,\theta)$ are the fragments with charge $Z_\text{true}$ and emission angle $\theta_\text{true}$ produced within the target in the angular acceptance of the TW detector.
The efficiencies are shown in Fig.~\ref{fig:eff} for each Z. 

As already mentioned the most important contribution to the efficiency reduction arises from the TW reconstruction.
This impact is expected to be more significant for the light fragments such as He and Li where more crowded events cause an increase of mismatch of the Z of two associated X-Y bars. For this reason events with mismatched Z are rejected in the analysis~\cite{FOOTGSI2019}.

\subsection{Systematic uncertainties}
\label{sec:sys}
Several sources of systematic uncertainties were identified in the analysis, both on the detectors involved and in the analysis method.
The systematic uncertainties coming from reconstruction at detector level were evaluated changing reconstruction or calibration strategy or parameters used in the reconstruction and propagating the variation through the full cross section analysis.
Finally the percentage variation bin by bin of the cross-section obtained with respect to the nominal one is assigned as uncertainty to the cross section measured with the data.
Two sources of systematics at detector level have been studied: the impact of the event selection performed using the BM and the requirements on the fragments reconstruction inside the TW.

The former systematic has been studied varying the selection criteria for the BM reconstructed tracks. A tight and a loose selection of the BM hits to be associated to a track, has been implemented and tested to verify the impact on the event pre-selection on the cross section measurements. The impact of this systematic uncertainty was found to be negligible.

Another source of systematic uncertainty is related to the charge reconstruction algorithm in the TW detector. This has been studied varying in the MC simulation the resolutions, within the experimental precision, in $\Delta$E, TOF and in the position measurements that affect the identification of the fragments in MC and thus the efficiencies and the purity evaluation. The only significant contribution was found to be the one related to the He ($<$~0.5$\%$). To check the impact on the ZID reconstruction also the Bethe-Bloch parametrized curves, extracted from a fit of the plot of $\Delta$E vs TOF at the true MC level, as explained in~\cite{FOOTGSI2019}, have been moved within the statistical uncertainty. The impact of this variation is small for all the fragments ($<$~0.5$\%$), and still maximum for Helium.
Finally the impact on the ZID coming from the TW calibration strategy was tested. In the TW calibration each $\Delta$E peak is fitted and calibrated to the MC value with a Birks fit~\cite{Morrocchi2019,Kraan2021,FOOTGSI2019}. Moving the $\Delta$E fitted mean value within the statistical uncertainty an impact which range in the interval~[0.1-2]\% has been found.

The systematic uncertainty on the unfolding procedure has been assessed by unfolding the MC angular distribution using a different unfolding method with respect to the nominal one (Bayesian iterative). The Iterative Dynamically Stabilized (IDS) method \cite{Malescu17} has been chosen as alternative method using the same number of iteration. The difference on the unfolded spectra between the two method range in the interval [0.1-3.6]\% where the biggest impact is on some angular bin of the heavier fragments.

Finally, the robustness of the reconstruction procedure, including the background subtraction exploiting target-out fragmentation, has been checked. The use of a background subtraction approach, given its statistical nature, introduced an uncertainty in the cross section reconstruction.
As reported in Sec.~\ref{sec:mc_corr}, the analysis method was validated looking at the agreement
between true and reconstructed MC cross sections after the unfolding procedure.
The difference between the true and the reconstructed MC cross sections takes into account all the intrinsic limitations of the adopted strategy.
This contribution was found to have an impact for all the fragments in the range [0.3-2.5]\% for total cross sections and it can be as high as $10\%$ for 
angular differential cross sections.

The numerical evaluation of the overall contributions of the systematic uncertainties is shown in Table~\ref{tab:xstotal} and in Table~\ref{tab:charge-angle-data}.

\section{Results}
\label{sec:results}
\begin{figure}
    \includegraphics[width=0.47\textwidth]{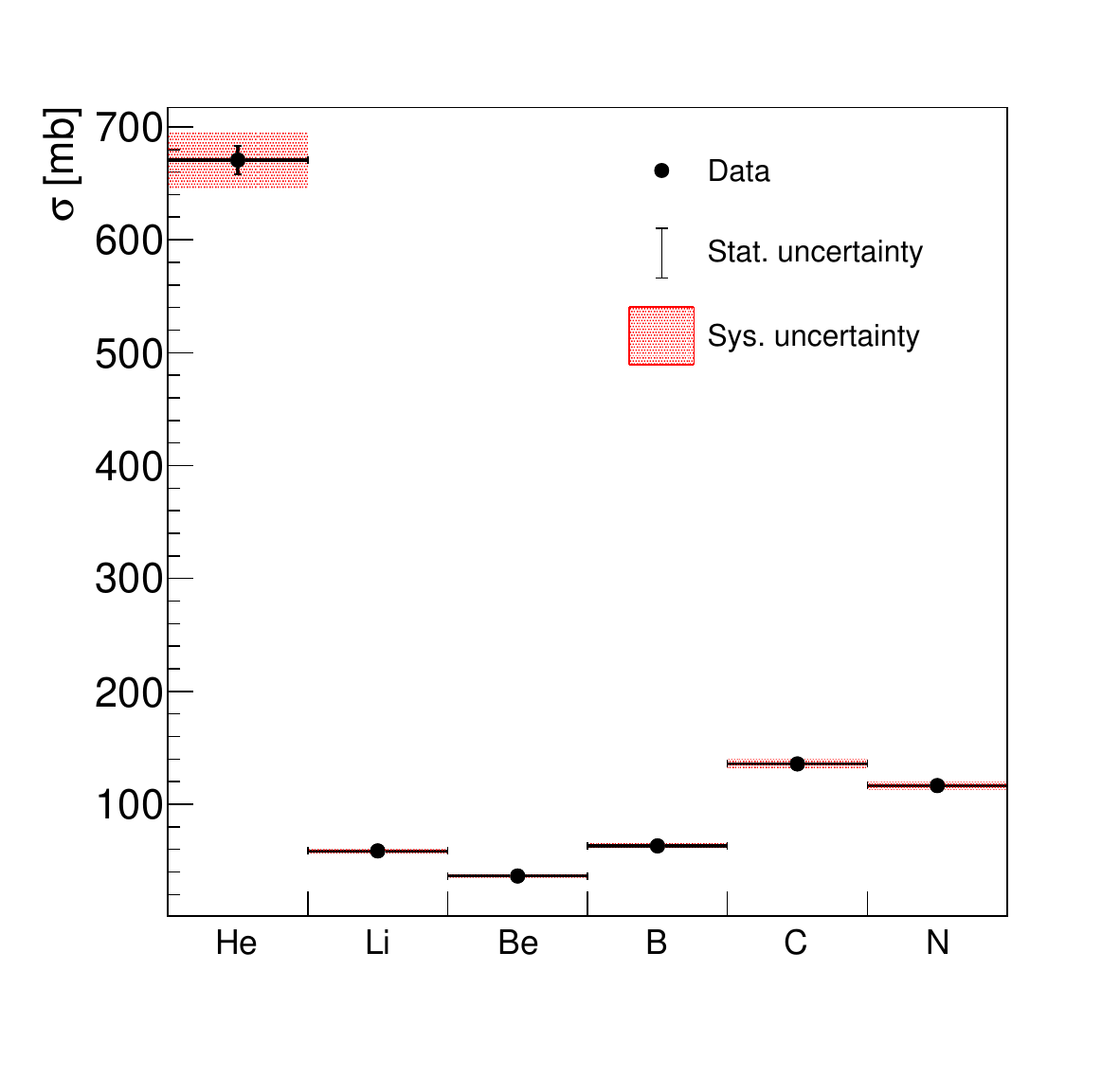}
    \caption{Elemental fragmentation cross sections for the production of He, Li, Be, B, C and N fragments in the interaction process of a \siO beam of 400 MeV/nucleon on a graphite target, integrated over the solid angle covered by the FOOT setup, $\theta\le 5.7^\circ$.}
    \label{fig:XStotal}
\end{figure}

In Fig.~\ref{fig:XStotal} the elemental fragmentation cross sections for the production of He, Li, Be, B, C and N in the interaction process of a \siO beam of 400 MeV/nucleon on a graphite target, integrated over the solid angle covered, $\theta\le 5.7^\circ$ are reported, along with statistic and systematic uncertainties (discussed in Sec.\ref{sec:sys}). Despite the \siO beam was delivered at an energy of 400\mev, the effective energy per
nucleon at the target was a bit lower due to previous energy losses and it was estimated by
MC simulation to be equal to 393\mev at the center of the target.
The cross sections have been measured according to Eq.\ref{eq:XStotal}, the yields of Eq.\ref{eq:YieldSub}, are considered after the unfolding procedure and all the needed corrections, discussed in Sec.\ref{sec:analysis}, have been applied. 

\begin{figure*}
    \centering
\begin{minipage}{0.45\textwidth}
  \centering
  \includegraphics[width=\textwidth]{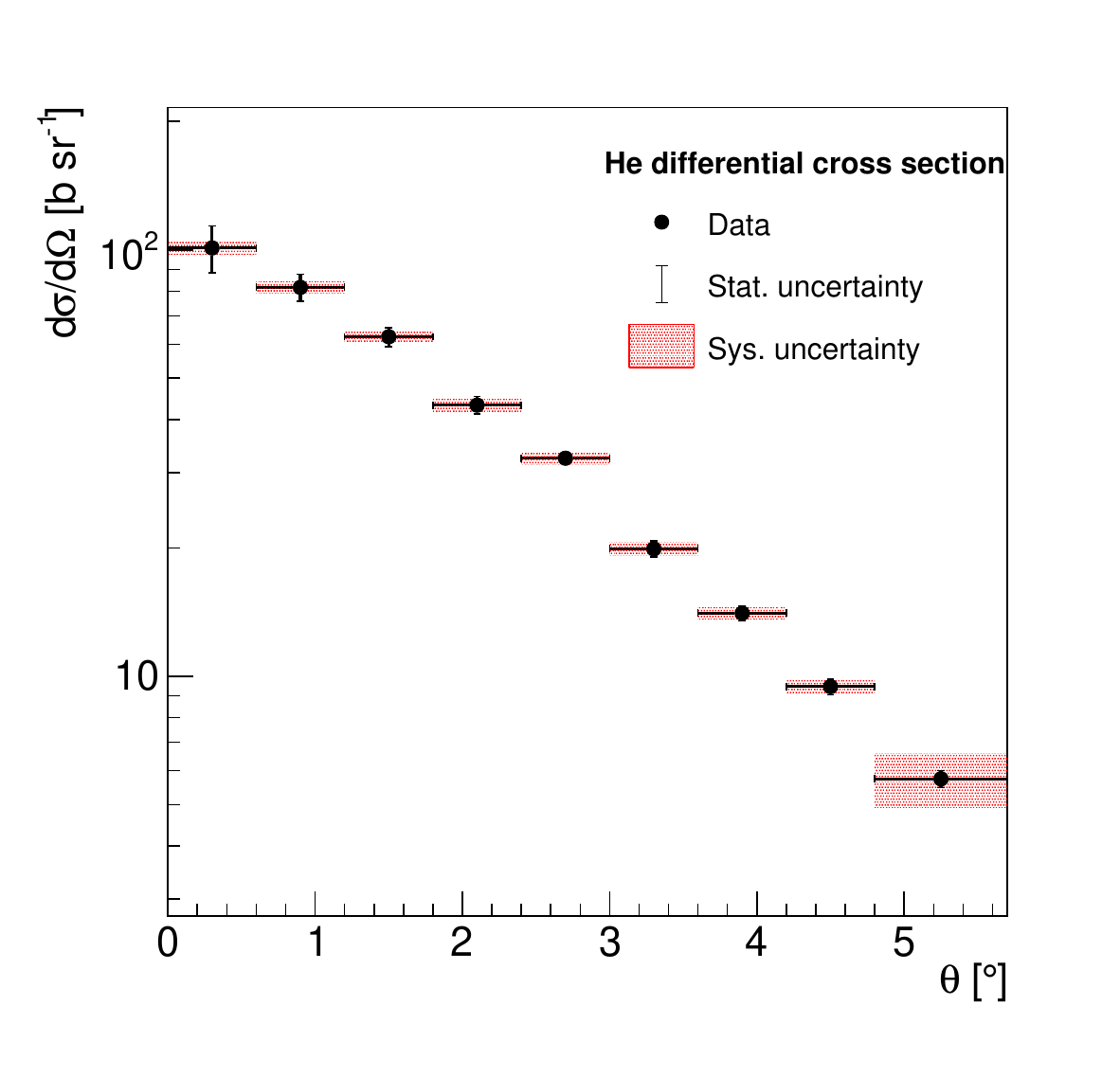}
  \\
  \includegraphics[width=\textwidth]{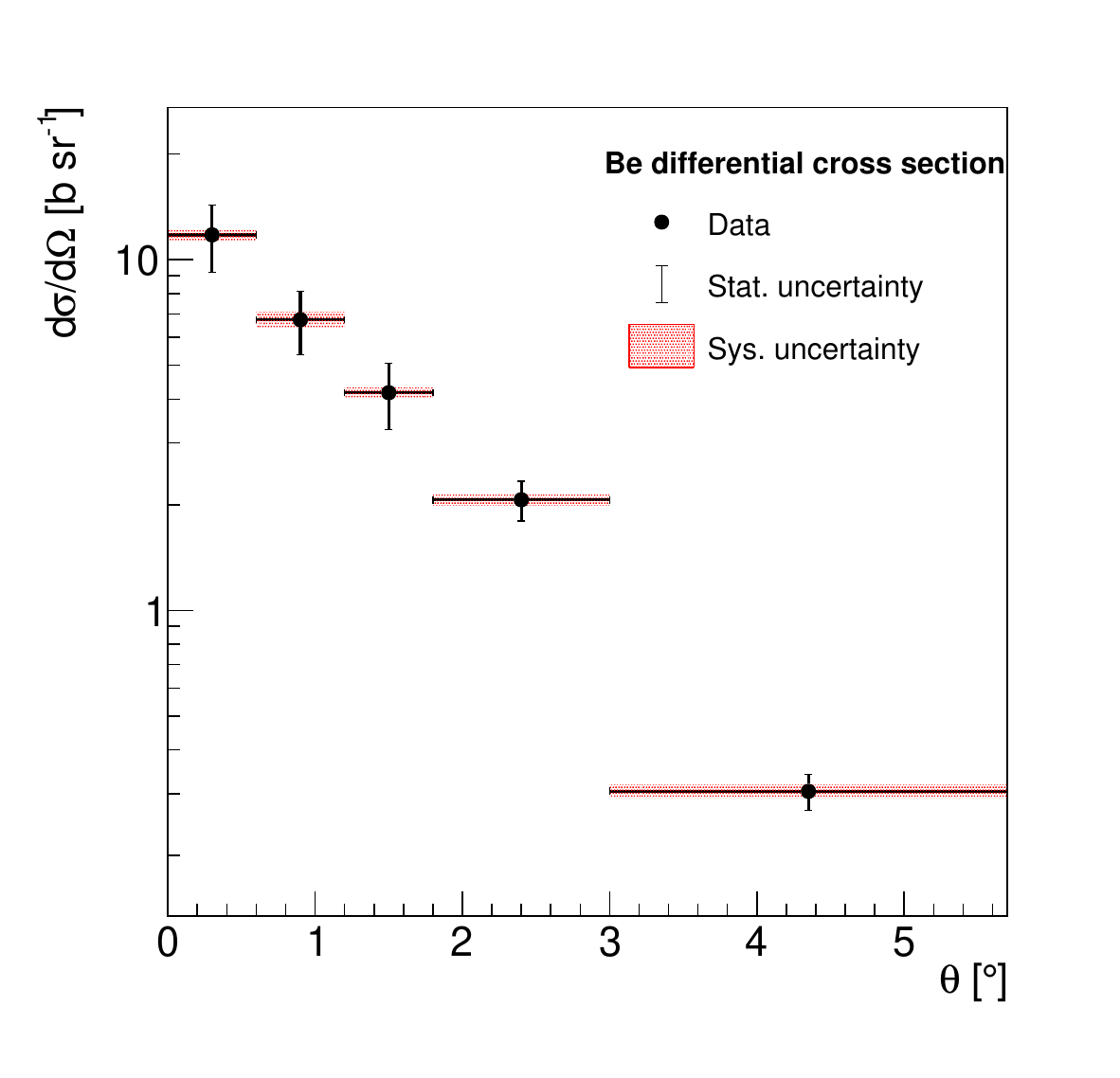}
  \\
   \includegraphics[width=\textwidth]{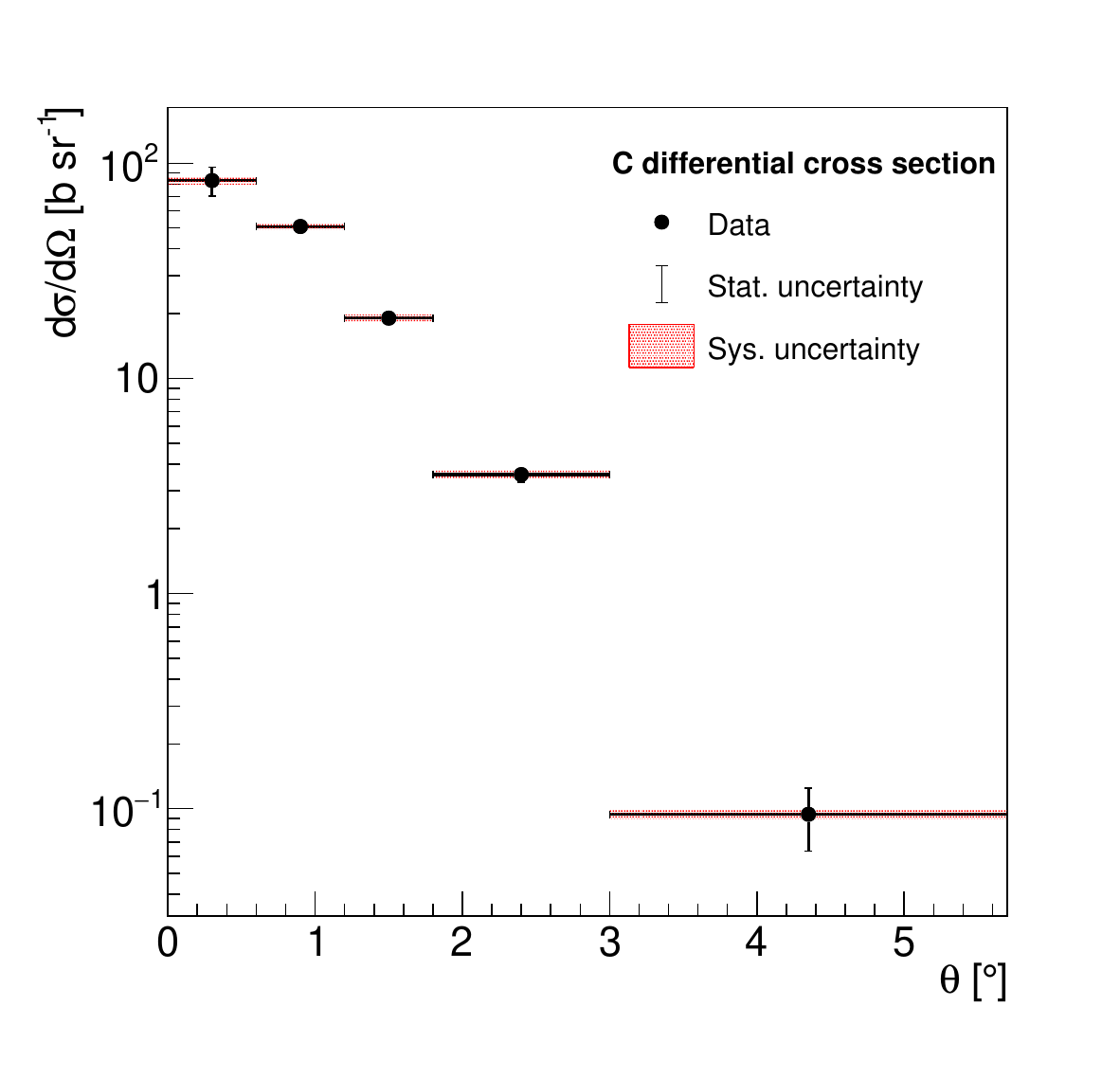}
\end{minipage}
\quad
\begin{minipage}{0.45\textwidth}
  \centering
  \includegraphics[width=\textwidth]{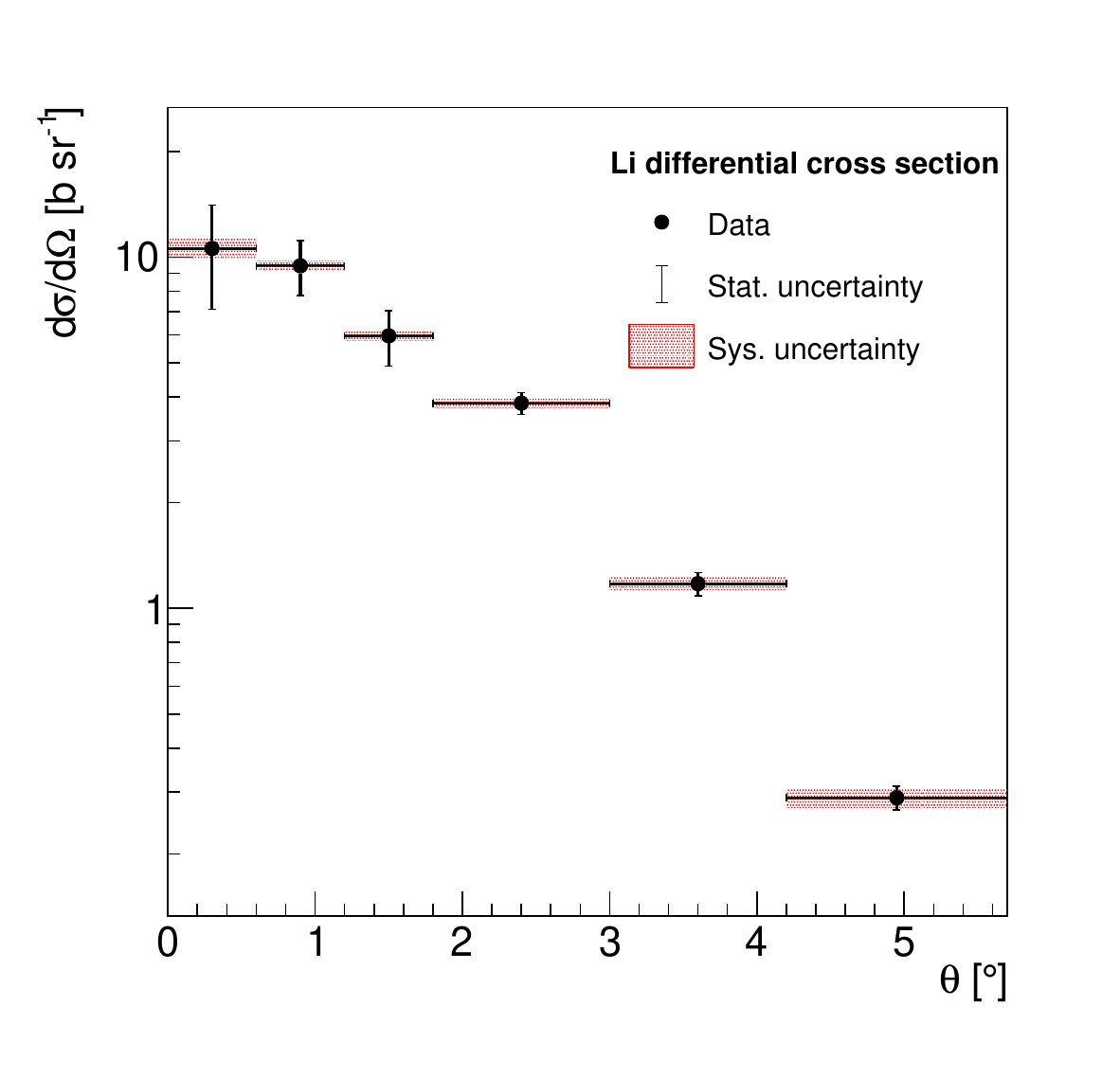}
  \\
  \includegraphics[width=\textwidth]{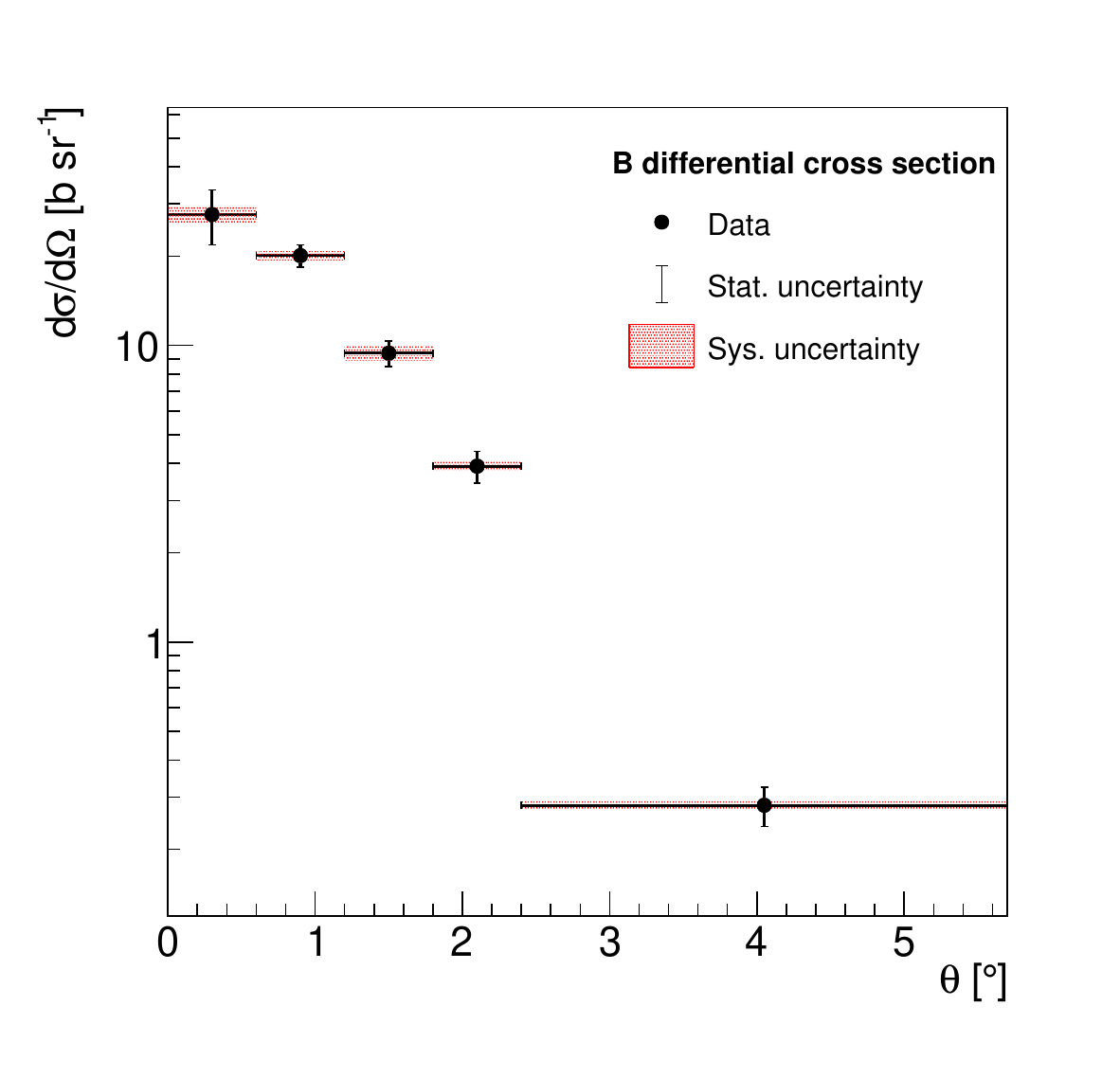}
  \\
  \includegraphics[width=\textwidth]{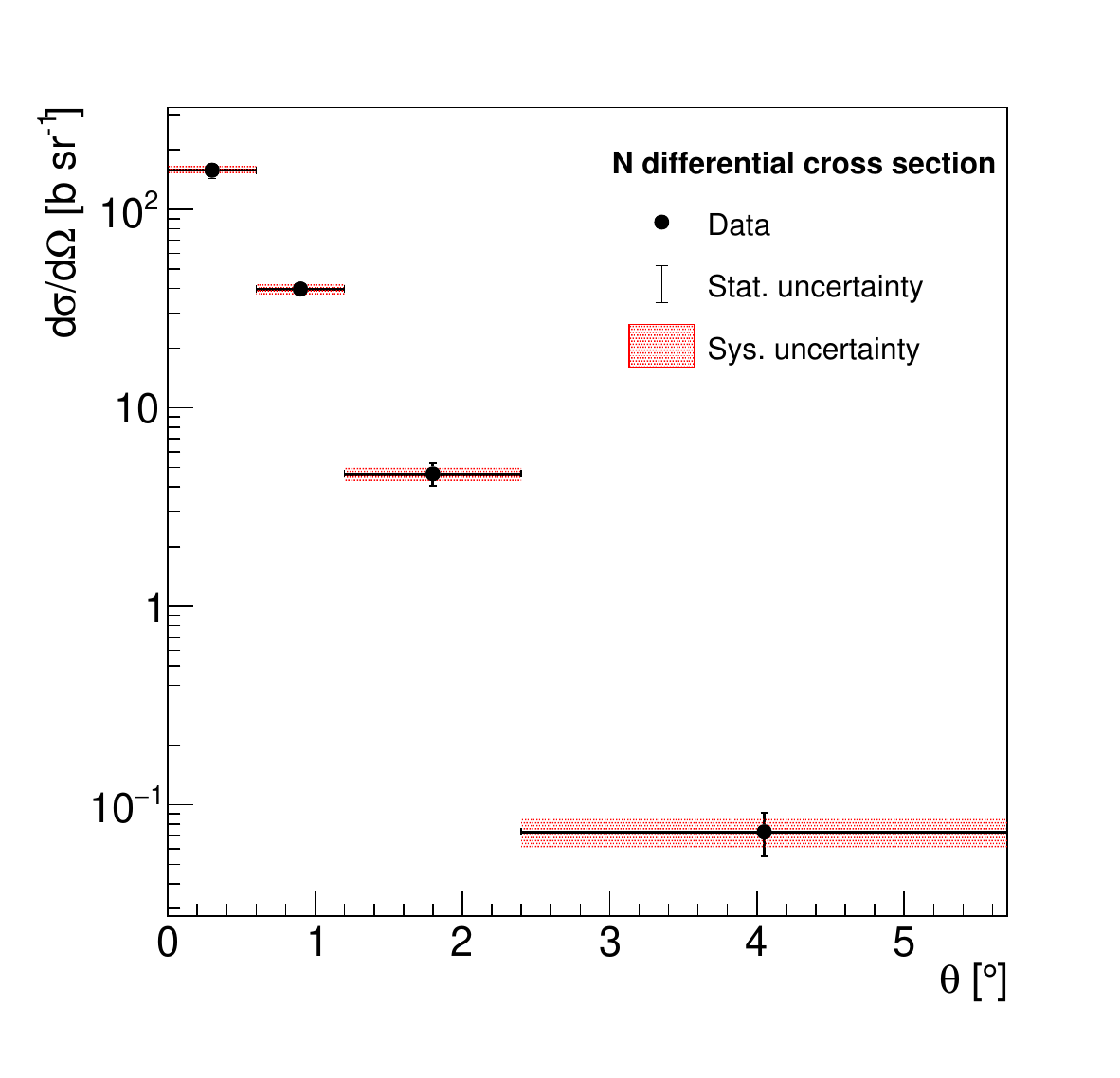}
\end{minipage}
\caption{Angular differential cross sections for the production of He, Li, Be, B, C and N fragments in the interaction process of a \siO beam of 400 MeV/nucleon on a graphite target. The results is given in bin of angle $\theta$ with $\theta\le 5.7^\circ$.}
\label{fig:XSangle}
\end{figure*}

The results for the elemental cross sections for He, Li, Be, B, C and N are also reported also in Table~\ref{tab:xstotal}, where the statistical and systematic uncertainties are reported separately and their weight on the
final value is reported. There it is possible to see that the systematic uncertainty
is always lower that the statistical one,
which is mainly driven by the limited statistics of the sample without target.
As previously reported from the 2019 FOOT campaign~\cite{FOOTGSI2019}, to our knowledge, there are no relevant measurements for He and Li at these energies, while there are some for $Z\geq 4$~\cite{Webber90,Zeitlin2011}. Moreover,
no new measurements were published between \cite{FOOTGSI2019} and the writing of this work.
The obtained elemental cross sections are then directly comparable with our previous measurements,
since the angular acceptance was the same and the velocity range of this work is slightly larger
but with a minimum impact on the results.
In our previous work~\cite{FOOTGSI2019}, the main comparison was performed with the work of Zeitlin and collaborators~\cite{Zeitlin2011}, a paper
providing elemental cross sections within an angular acceptance of $\simeq 7^{\circ}$ with a
375\mev \siO beam, an energy slightly lower with respect to this work.
In particular, in~\cite{Zeitlin2011}, only $Z \geq 5$ cross sections were provided, allowing a fair comparison despite the different angular acceptance given the mostly forward production of such fragments.
In our previous work~\cite{FOOTGSI2019} we concluded that the results seemed to confirm those in \cite{Zeitlin2011}.
In this work, we confirm our previous measurements regarding Be, C and N while for He, B this result is slightly lower but still comparable within the uncertainties. The largest difference involves the Li results.
This is due to the impact of the purity correction (see Sec.~\ref{sec:purity}) which was completely underestimated in the previous work~\cite{FOOTGSI2019}. The observed discrepancies between the two analysis come from this correction, together with the very few statistics collected in that campaign.
The effect of the absence of the purity correction in previous work is the one of overestimating the cross sections for those fragments for which the correction is larger. While the correction is almost negligible for most of the fragments, for Li is important due to the events of He ions pairs which release in a TW bars' crossing an energy similar to the one released by a Li, as discussed in Sec.~\ref{sec:purity}.

The final results for the angular differential cross sections are shown in Fig.~\ref{fig:XSangle} and in Table~\ref{tab:charge-angle-data}. To our knowledge, in this case, there are no previous results to
compare with. Thanks to the unfolding techniques, we succeeded in reducing the systematic uncertainties
on the angular spectrum although the limited statistics of the without target sample gave an important
contribution to the statistical uncertainties going from $3\%$ to $20\%$ on average except for
a $33\%$ contribution in the first bin of Li.
As already mentioned, the number of bins and their width were carefully chosen considering
the available statistics in order to have a reasonable number of fragments in each bin.

\begin{table}
\begin{ruledtabular}
\begin{tabular}{cccc}
\textrm{Element}&
$\sigma\pm\Delta_{stat}\pm\Delta_{sys}\,[\textrm{mb}]$&
$\Delta_{stat}/\sigma$&
$\Delta_{sys}/\sigma$\\
\colrule
\textrm{He} & $671 \pm 12 \pm 25$ & $1.9\%$ & $3.7\%$ \\
\textrm{Li} & $59 \pm 3 \pm 2$ & $5.6\%$ & $3.5\%$  \\
\textrm{Be} & $37 \pm 3 \pm 1$ & $7.8\%$ & $3.0\%$  \\
\textrm{B} & $63 \pm 4 \pm 2$ & $6\%$ & $3\%$  \\
\textrm{C} & $136 \pm 6 \pm 4$ & $4.4\%$ & $3.1\%$  \\
\textrm{N} & $117 \pm 6 \pm 4$ & $5.4\%$ & $3.0\%$  \\
\end{tabular}
\end{ruledtabular}
\caption{Elemental cross sections measured in this work. The contribution of the statistical and systematic uncertainties is reported separately.}
\label{tab:xstotal}
\end{table}

\begin{table}[h]
\centering
\begin{tabular}{ccccc}
\hline
Z & $\theta [^{\circ}]$ & $d\sigma/d\Omega\pm\Delta_{stat}\pm\Delta_{sys}$ & $\Delta_{stat}^{rel}$ & $\Delta_{sys}^{rel}$ \\
 & & $[\textrm{b sr}^{-1}]$ &  & \\ \hline

\multirow{9}{*}{He} & $0-0.6$ & $101\pm 13 \pm 3$ & $12.6\%$ & $3.4\%$ \\
 & $0.6-1.2$ & $82 \pm 6 \pm 3$ & $7.3\%$ & $3.1\%$\\
 & $1.2-1.8$ & $63 \pm 3 \pm 2$ & $5.1\%$ & $3\%$\\
 & $1.8-2.4$ & $43.2 \pm 2.1 \pm 1.5 $ & $4.8\%$ & $3.4\%$\\ 
 & $2.4-3$& $32.5 \pm 1.2 \pm 1.1$ & $3.6\%$ & $3.2\%$\\ 
 & $3-3.6$& $19.9 \pm 0.9 \pm 0.7$ & $4.4\%$ & $3.4\%$\\ 
 & $3.6-4.2$& $14.1\pm 0.6 \pm 0.4$ & $4.2\%$ & $3.1\%$\\ 
 & $4.2-4.8$& $9.5\pm 0.4 \pm 0.3$ & $4.2\%$ & $3.1\%$\\ 
 & $4.8-5.7$& $5.7\pm 0.3 \pm 0.8$ & $4.6\%$ & $14.5\%$\\ \hline
\multirow{6}{*}{Li} & $0-0.6$ & $10.6\pm 3.5 \pm 0.6$ & $32.9\%$ & $6.0\%$ \\
 & $0.6-1.2$ & $ 9.4\pm 1.7 \pm 0.3$ & $17.9\%$ & $3.2\%$\\
 & $1.2-1.8$ & $ 6.0\pm 1.1 \pm 0.2$ & $18.0\%$ & $3.0\%$\\
 & $1.8-3$ & $ 3.8\pm 0.3 \pm 0.1 $ & $7.2\%$ & $3.2\%$\\ 
 & $3-4.2$& $ 1.18\pm 0.09 \pm 0.05$ & $7.6\%$ & $4.1\%$\\
 & $4.2-5.7$& $ 0.289\pm 0.022 \pm 0.018$ & $7.8\%$ & $6.2\%$\\ \hline
\multirow{5}{*}{Be} & $0-0.6$ & $11.8\pm 2.6 \pm 0.4$ & $22\%$ & $3\%$ \\
 & $0.6-1.2$ & $ 6.7\pm 1.4 \pm 0.3$ & $21\%$ & $5\%$\\
 & $1.2-1.8$ & $ 4.2\pm 0.9 \pm 0.1$ & $21.4\%$ & $3.2\%$\\
 & $1.8-3$ & $ 2.1\pm 0.3 \pm 0.1$ & $13\%$ & $3\%$\\
 & $3-5.7$ & $ 0.305\pm 0.036 \pm 0.013$ & $12\%$ & $4\%$\\ \hline
\multirow{5}{*}{B} & $0-0.6$ & $27.6\pm 5.8 \pm 1.7$ & $21\%$ & $6\%$ \\
 & $0.6-1.2$ & $20.1 \pm 1.7 \pm 0.6$ & $9\%$ & $3\%$\\
 & $1.2-1.8$ & $9.4 \pm 0.9 \pm 0.5$ & $9.8\%$ & $5.2\%$\\
 & $1.8-2.4$ & $3.9 \pm 0.5 \pm 0.1$ & $12\%$ & $3\%$\\
 & $2.4-5.7$ & $0.282 \pm 0.043 \pm 0.009$ & $15\%$ & $3\%$\\ \hline
 \multirow{5}{*}{C} & $0-0.6$ & $83\pm 13 \pm 3$ & $15\%$ & $4\%$ \\
 & $0.6-1.2$ & $ 51\pm 3 \pm 2$ & $5.7\%$ & $3.4\%$\\
 & $1.2-1.8$ & $ 19.1\pm 1.3 \pm 0.7$ & $6.9\%$ & $3.9\%$\\
 & $1.8-3$ & $ 3.6\pm 0.3 \pm 0.2$ & $8.0\%$ & $4.5\%$\\
 & $3-5.7$ & $ 0.09\pm 0.03 \pm 0.01$ & $32.6\%$ & $5.3\%$\\ \hline
 \multirow{4}{*}{N} & $0-0.6$ & $158\pm 14 \pm 7$ & $9\%$ & $4\%$ \\
 & $0.6-1.2$ & $ 39.7\pm 2.8 \pm 2.7$ & $7.1\%$ & $6.8\%$\\
 & $1.2-2.4$ & $ 4.6\pm 0.6 \pm 0.4$ & $13\%$ & $8\%$\\
 & $2.4-5.7$ & $ 0.073\pm 0.018 \pm 0.012$ & $25\%$ & $17\%$\\ \hline

 \end{tabular}
\caption{Angular differential cross section measured in this work. The contribution of the statistical and systematic uncertainties is reported separately.}
\label{tab:charge-angle-data}
\end{table}

\section{Comparison with prominent nuclear interaction models}
\label{sec:MCcomparison}

\begin{figure*}
    \centering
\begin{minipage}{0.45\textwidth}
  \centering
  \includegraphics[width=\textwidth]{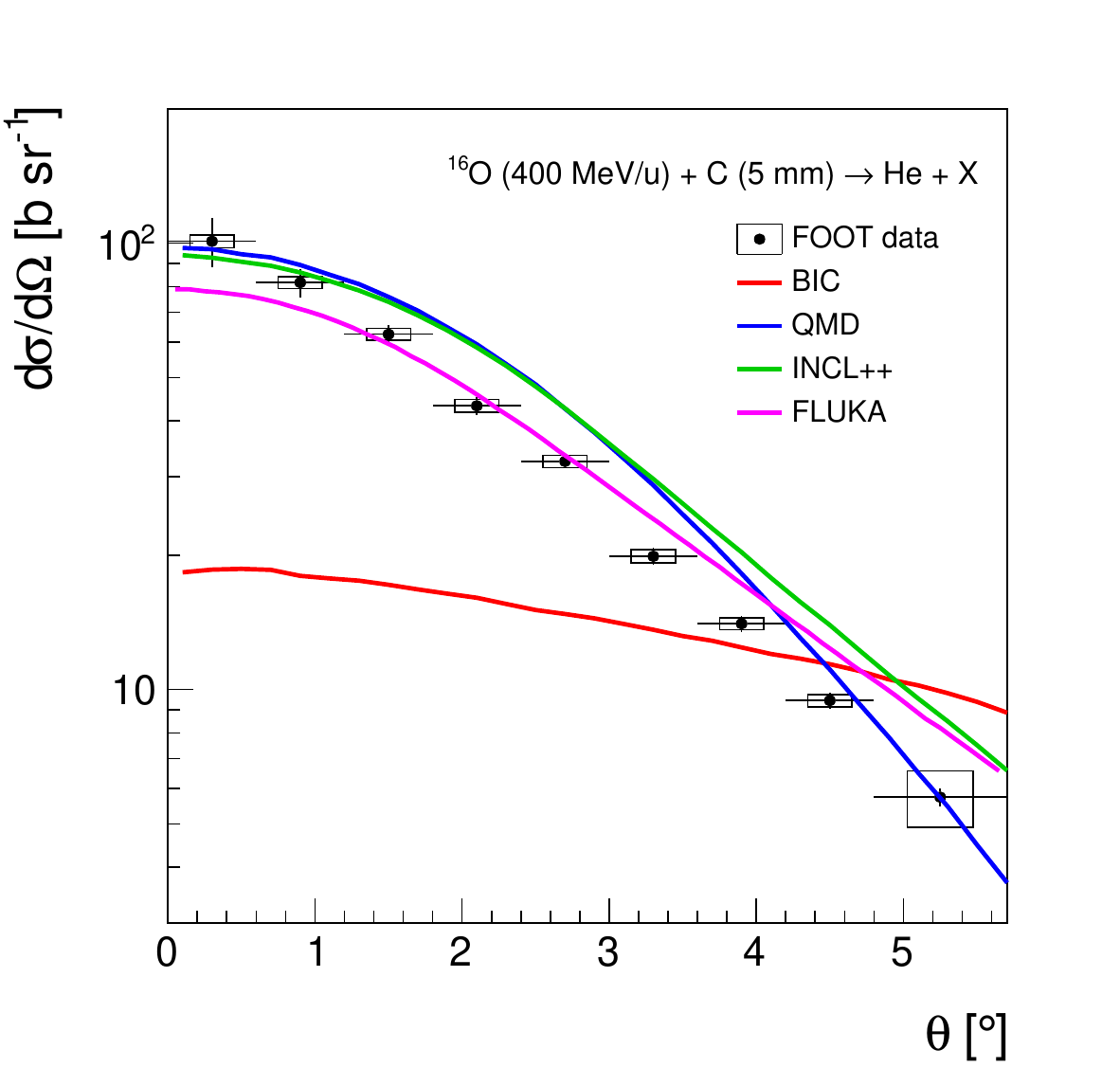}
  \\
  \includegraphics[width=\textwidth]{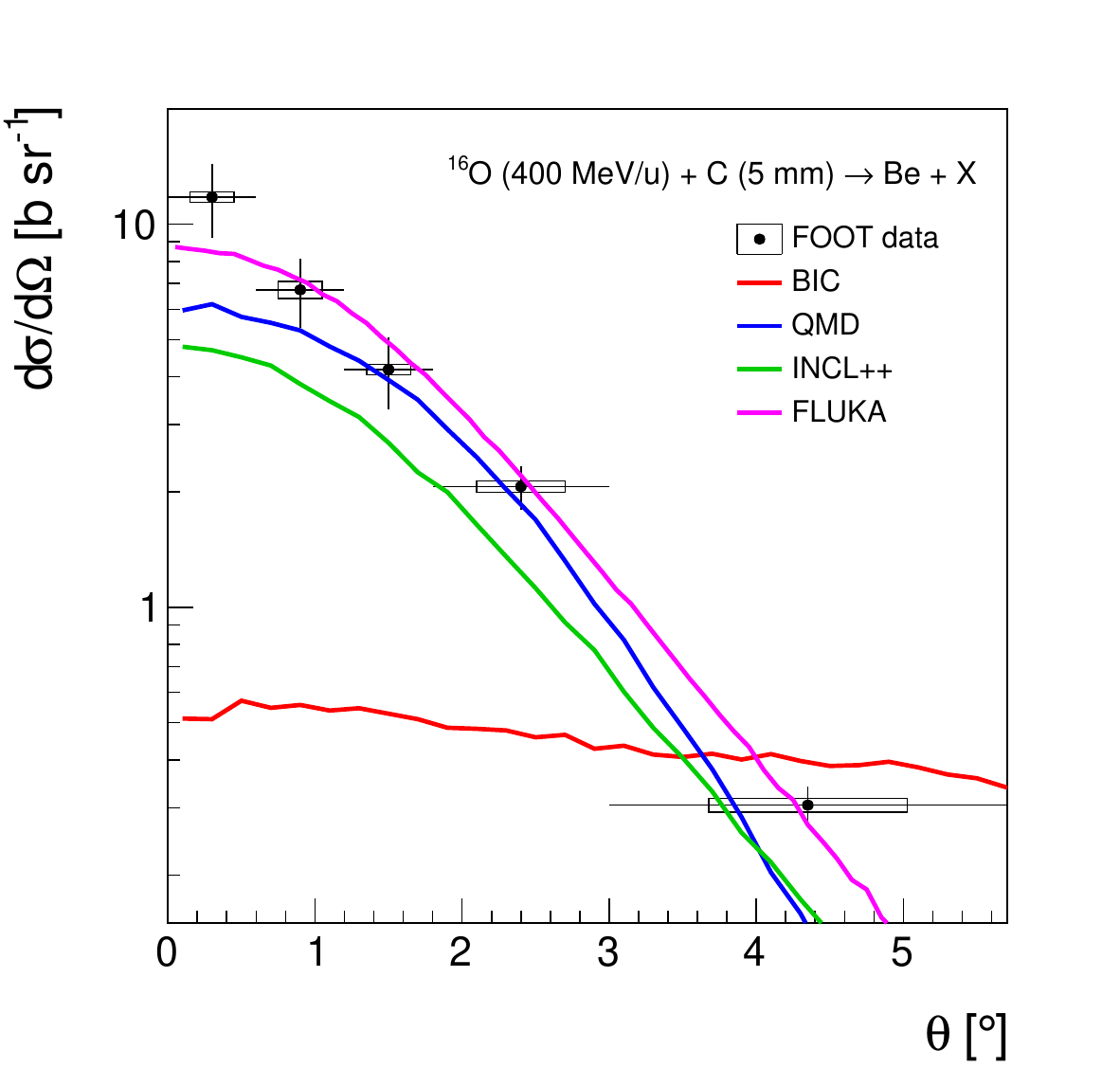}
  \\
  \includegraphics[width=\textwidth]{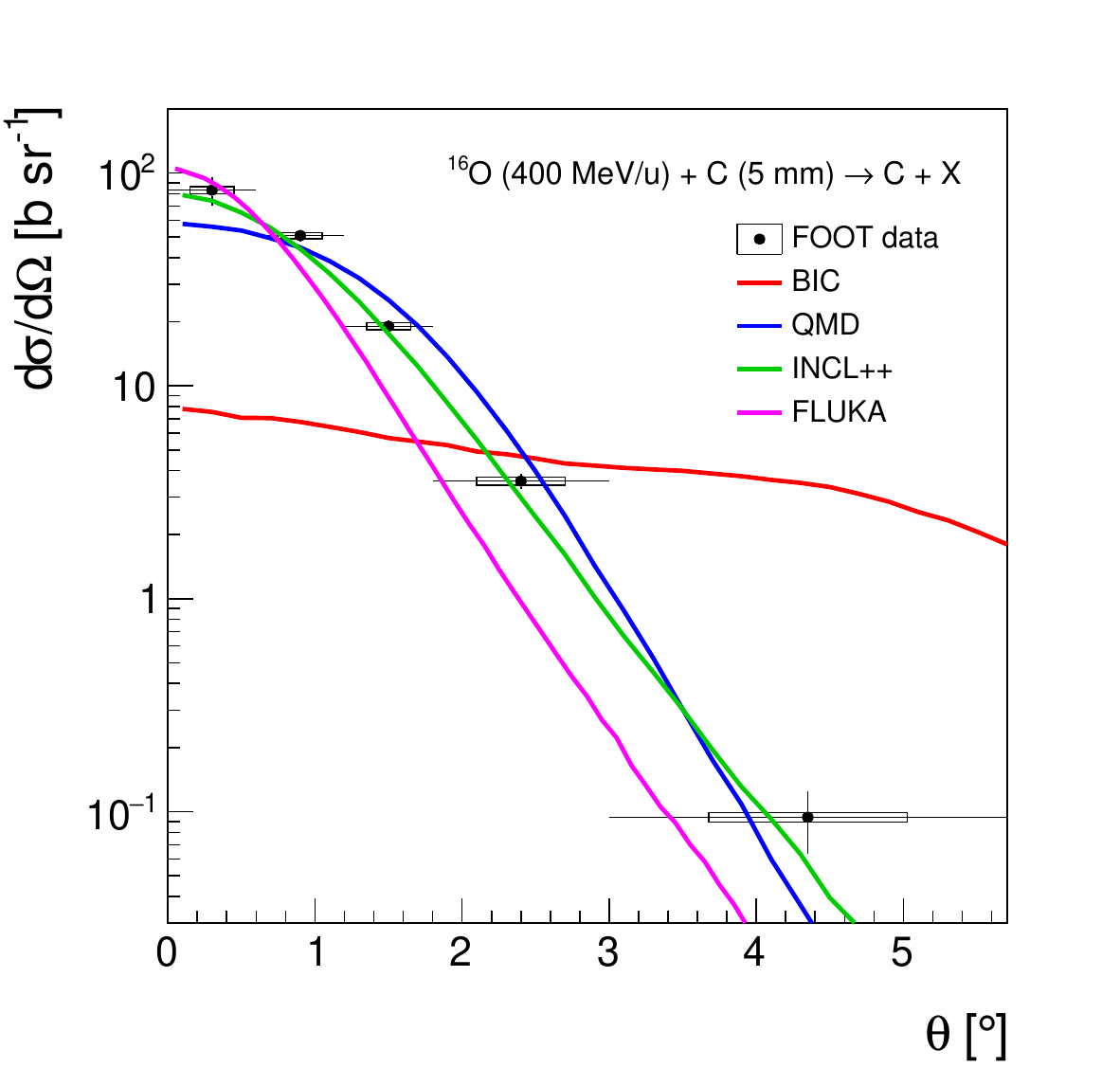}
\end{minipage}
\quad
\begin{minipage}{0.45\textwidth}
  \centering
  \includegraphics[width=\textwidth]{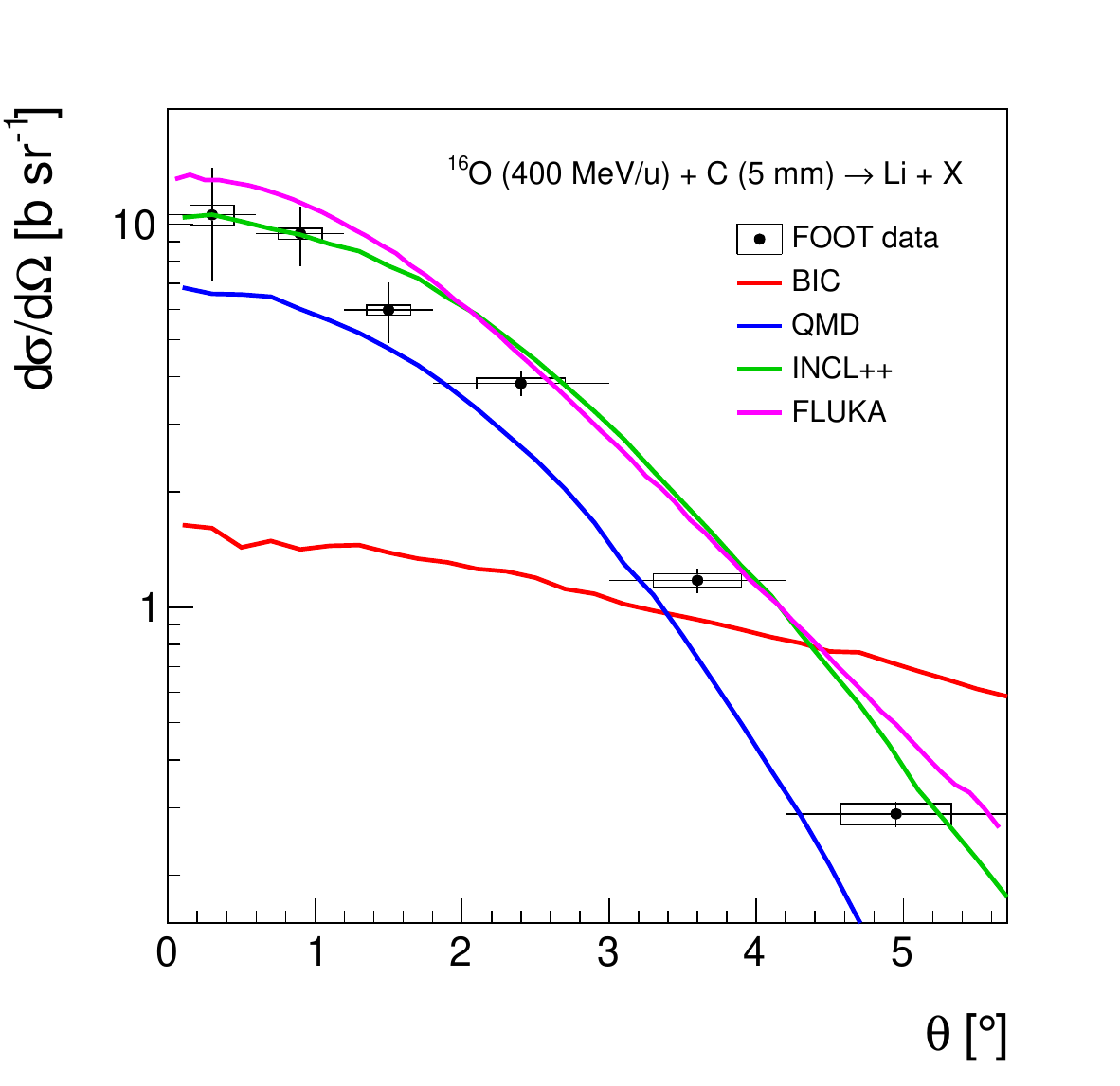}
  \\
  \includegraphics[width=\textwidth]{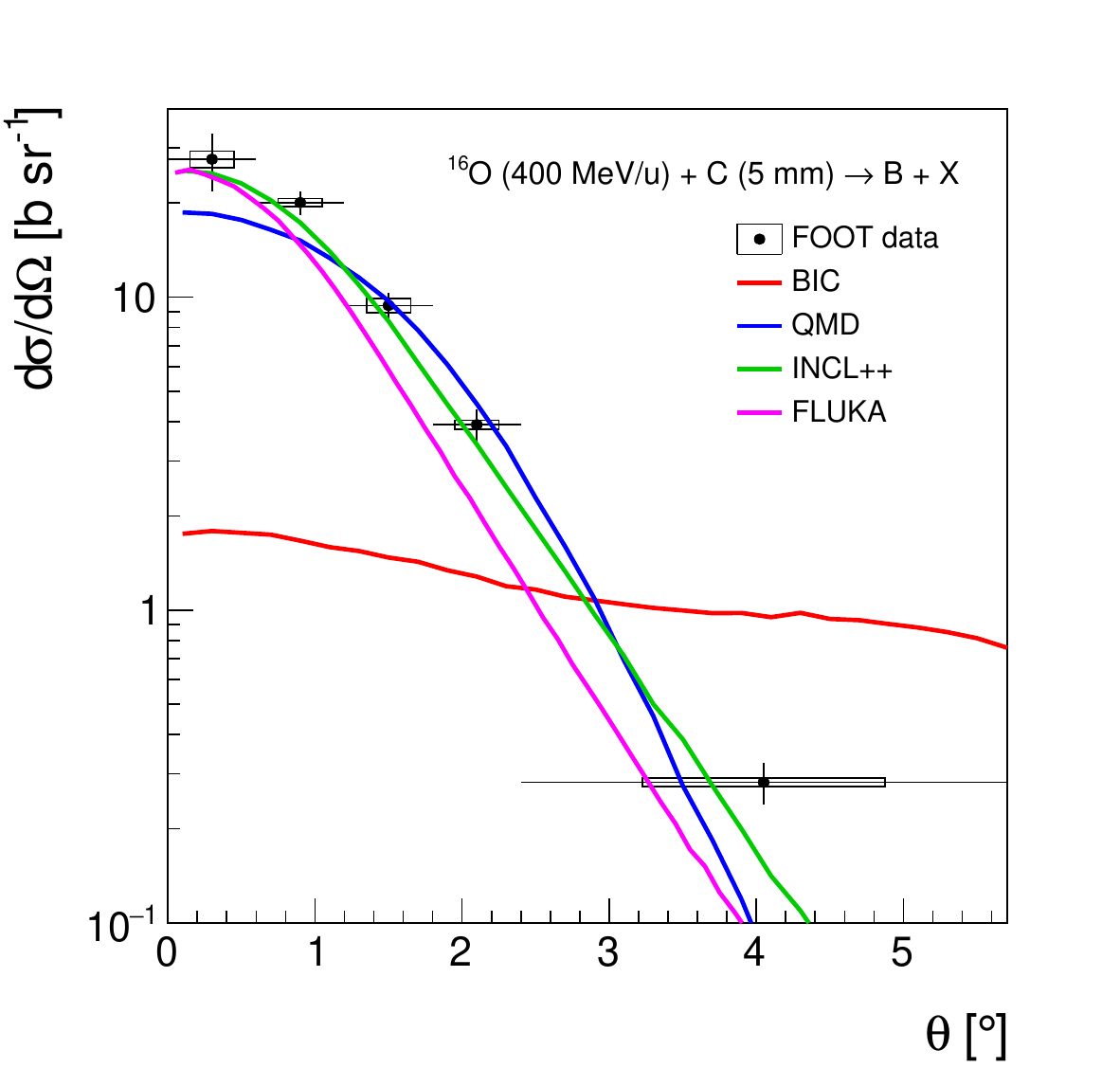}
  \\
  \includegraphics[width=\textwidth]{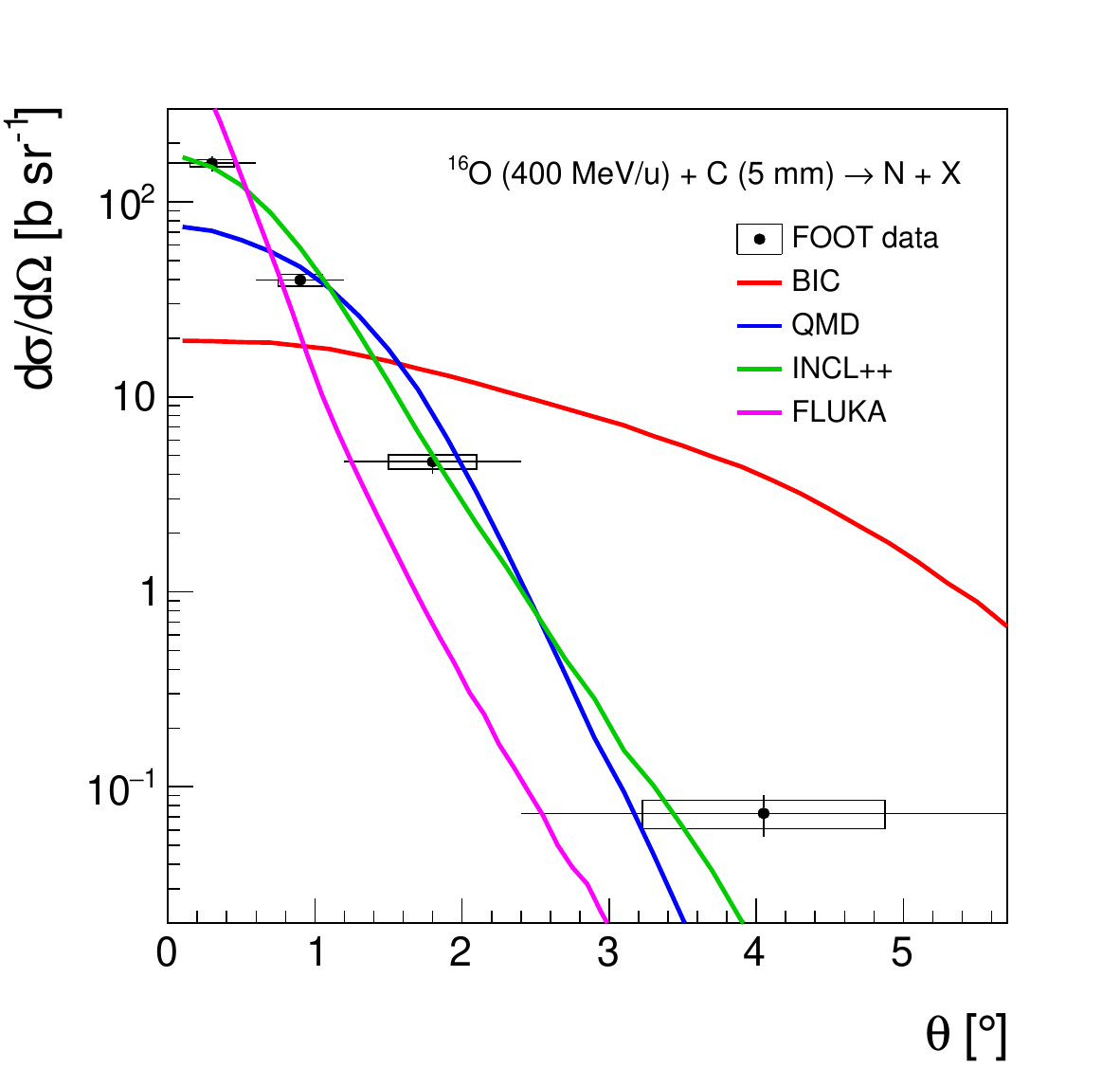}
\end{minipage}
\caption{Angular differential cross sections for the production of He, Li, Be, B, C and N fragments in the interaction process of a \siO beam of 400 MeV/nucleon on a graphite target together with FLUKA and Geant4 predictions with four different models (see insert).}
\label{fig:XSMCcomparison}
\end{figure*}

The experimental angular differential cross sections were compared with theoretical predictions 
from four different nuclear interaction models used in Monte Carlo codes, as shown in Fig.~\ref{fig:XSMCcomparison}. FLUKA \cite{Battistoni2016} was used alongside three distinct Geant4 \cite{Agostinelli2003,Allison2006,Allison2016} hadronic models: Binary Ion Cascade (BIC) \cite{Folger2004}, Quantum 
Molecular Dynamics (QMD) \cite{Mancusi2009}, and Liège Intranuclear Cascade (INCL$^{++}$) \cite{Mancusi2014}. 

BIC was implemented using G4IonPhysics, a specialized physics library designed to handle ion-ion 
interactions. It implements a combination of models to cover a wide range of energies and 
interaction types for light ions (deuteron, triton, $^3$He, alpha) and generic ions ($Z>2$).
The G4IonPhysics library uses Glauber-Gribov cross sections for all nucleus-nucleus interactions implementing the Binary Light Ion (BIC) model with Precompound/de-excitation for the
low energy regime and the FTFP (Fritiof string model with Precompound/de-excitation) for energies
above $3\,\text{GeV/nucleon}$.

QMD was implemented via G4LightIonQMDPhysics library which is often used for applications involving light ion transport, such as medical physics and space radiation studies.
This library provides a microscopic approach to simulate nuclear collisions by treating nucleons as wave packets that evolve according to quantum molecular dynamics. This approach allows for a more realistic representation of nuclear interactions compared to simplified models.

Both these models are extended with the use of G4HadronPhysicsQGSP\_BIC which represents a comprehensive approach to hadronic interaction modeling, combining the Quark-Gluon String Precompound (QGSP) model with the Binary Intranuclear Cascade (BIC) model. This physics library finds wide application in various simulation scenarios, from high-energy physics experiments to medical applications.

The third model used in Geant4 is INCL$^{++}$ which treats nuclear reactions as a series of independent nucleon-nucleon collisions within the target nucleus.
It models nucleons as a free Fermi gas within a static potential well. It defines participants as 
projectile nucleons entering a calculated target volume. The collision is treated asymmetrically: 
a quasi-projectile is formed from projectile spectators and non-cascading participants, while the 
quasi-target final state results from the full cascade dynamics within the calculation volume, 
yielding a more reliable description of the target remnant. Due to its strength in reproducing 
target fragmentation, INCL$^{++}$ typically operates in inverse kinematics (target impacting 
projectile) by default to better simulate projectile fragmentation. If both projectile and target
mass number exceed $A=18$, the simulation defaults to using the aforementioned BIC model instead.

Regarding FLUKA simulation, for projectile energies above $150\,\text{MeV/nucleon}$ nucleus-nucleus collisions are treated using an interface to a modified rQMD-2.4 (relativistic Quantum Molecular Dynamics) model \cite{Sorge1989,Sorge1995}, which can also be run as intranuclear cascade. Examples of FLUKA results compared to experimental data when running the modified rQMD-2.4 model can be found in \cite{Andersen2004,Aiginger2005,Ballarini2007}.
Although initially developed for high energies, the model could be extended down to the energies of interest of the FOOT experiment. The model is also coupled to a pre-equilibrium stage, which, in FLUKA, is managed by the PEANUT (PreEquilibrium Approach to Nuclear Thermalization) model \cite{Ferrari19977,Ferrari2002}. Here pre-equilibrium is based on the developments of the GDH (Geometry Dependent Hybrid model) exciton model \cite{Blann1971}. The late stages of the interaction (fragmentation and evaporation) are then modeled, for nuclei with $A<17$, by means of a phase space Fermi Break-up model \cite{Fermi1950,Epherre1967}.

Across all six plots of Fig.~\ref{fig:XSMCcomparison}, the BIC model consistently and significantly underestimates the experimental cross section data predicting a much flatter angular distribution than the data, especially for the lighter fragments.
Regarding FLUKA simulation, for He, Li, Be and B it shows generally good agreement with the data. The model used in FLUKA is able to capture the trend of decreasing cross section with increasing angle quite well apart from some deviations, for instance predicting a narrower spectrum for B
at larger angles while for C and N fragments it tends to underpredict the value of the cross section.
The INCL$^{++}$ model shows very good agreement with the data for most fragments following experimental points across the whole angular range except for a consistent underestimation of
Be cross section.
The QMD gives also similar results but it appears to underestimate the cross sections at certain angles predicting a narrower angular distribution for most fragments, in particular for Li and Be.
In conclusion, FLUKA, INCL$^{++}$, and QMD models show significantly better agreement with the FOOT experimental data compared to the BIC model, which consistently underestimates the cross sections. Among the better-performing models, INCL$^{++}$ and QMD appear to provide the most consistently accurate predictions across the range of fragments (from He to N) shown in these plots, closely followed by FLUKA, which performs well for lighter fragments but tends to underestimate the production of heavier fragments like C and N in this specific reaction and energy regime.

\section{Conclusions}
This paper presents the analysis for the measurement of the elemental fragmentation cross sections and the first measurement of differential angular cross sections for the forward production ($\theta\leq 5.7 ^\circ$) of He, Li, Be, B, C and N nuclei by a 400\mev \siO beam interacting with a 5~mm graphite target. 
The integrated cross sections were already published by the FOOT collaboration~\cite{FOOTGSI2019} and also in other works ~\cite{Zeitlin2011,Webber90} only for the production of fragments with Z$\ge$5.
A reduced FOOT setup, consisting of a beam monitor drift chamber and a system of scintillating detectors for the energy loss ($\Delta$E) and the time of flight (TOF) measurements, provided sufficient identification capabilities to resolve fragments with different charge Z and measure their emission angles.
As with previous measurements~\cite{FOOTGSI2019} the overall relative statistical uncertainty exceeds the systematic uncertainties, indicating that limited statistics significantly impact measurement precision. In particular the low statistics of the sample without the target is primarily responsible of the overall relative statistical uncertainty. This limitation also affected the number of bins chosen for the differential angular cross sections, requiring a compromise between available statistics and the number of bins.
Despite the reduced experimental setup and available statistics, the measured elemental fragmentation cross sections successfully addressed the data gaps needed for PT and RPS applications, particularly regarding angular differential cross sections measured for the first time.
The integrated cross section results can be compared to those previously published by our collaboration~\cite{FOOTGSI2019} and by Zeitlin and collaborators~\cite{Zeitlin2011}.
This analysis confirmed our previous measurements of Be, C and N cross sections while for He and B the results are slightly lower but still comparable within uncertainties. The differences observed for these ions, and particularly for Li, are due to the impact of the purity correction not considered in the previous work. For this reason, these new results supersede the ones presented in~\cite{FOOTGSI2019}. Our results are in agreement with the results in~\cite{Zeitlin2011} where they measured the elemental cross section production of B, C and N, with a setup with an angular acceptance similar to the one used for this analysis and for a \siO energy beam of similar energy. As stated in~\cite{FOOTGSI2019}, these results helped discriminate between conflicting results from earlier studies~\cite{Webber90,Zeitlin2011}.
The angular differential cross sections were compared with theoretical predictions 
from prominent nuclear interaction models for the energy range and interacting systems of interest for FOOT. 
While the BIC model fails to accurately reproduce both the magnitude and angular dependence of the cross sections, the FLUKA model, as well as the QMD and INCL$^{++}$ models from Geant4, show good agreement with the experimental data. Among these, the INCL$^{++}$ model most closely matches the measured angular distributions and cross section values, with the notable exception of the Beryllium (Be) data.
The performed measurements demonstrate that the FOOT experimental setup operated in the 2021 GSI campaign, even with a limited number of collected events and with a reduced number of operating detectors, can provide valuable cross section measurements.

\section*{Acknowledgments}
We thank GSI for the successful operation of their facilities during the data taking. The FOOT Collaboration acknowledges the INFN for its support in building and running the detector. We would like to acknowledge all the personnel of the CNAO and GSI centres that provided us support during the operational tests performed using proton, \twC and \siO beams at their facilities.

\bibliography{gsi}
\end{document}